\documentclass[number,5p]{elsarticle}

\usepackage[utf8]{inputenc}
\usepackage[T1]{fontenc}
\usepackage{graphicx}
\usepackage{grffile}
\usepackage{longtable}
\usepackage{wrapfig}
\usepackage{rotating}
\usepackage[normalem]{ulem}
\usepackage{amsmath}
\usepackage{textcomp}
\usepackage{amssymb}
\usepackage{capt-of}
\usepackage{hyperref}
\usepackage{wasysym}
\usepackage{gensymb}
\usepackage{hepnames}
\usepackage[inline]{enumitem}

\usepackage[per-mode=symbol, range-units=single, binary-units=true]{siunitx}
\DeclareSIUnit\clight{\text{\ensuremath{c}}} 
\DeclareSIUnit\momentum{\GeV\per\clight} 
\DeclareSIUnit\tmom{(\momentum)^2} 
\DeclareSIUnit\atom{\text{atoms}} 
\DeclareSIUnit\event{\text{events}} 
\DeclareSIUnit\KB{\text{KB}} 

\usepackage{booktabs}
\newcommand\headercell[1]{%
\smash[b]{\begin{tabular}[t]{@{}c@{}} #1 \end{tabular}}}

\usepackage{ragged2e}

\begin{document}

\begin{frontmatter}	
  \title{The KOALA Experiment for (anti)proton-proton Elastic Scattering}
  \date{\today}

  \author[ikp]{Yong Zhou\corref{cor}}
  \ead{y.zhou@fz-juelich.de}
  \author[ikp]{Huagen Xu}

  \author[ikp]{Ulf Bechstedt}
  \author[ikp]{Jürgen Böker}
  \author[ikp]{Nils Demary}
  \author[ikp]{Frank Goldenbaum}
  \author[ikp]{Dieter Grzonka}
  \author[ikp]{Jan Hetzel}
  \author[muenster]{Alfons Khoukaz}
  \author[ikp]{Franz Klehr}
  \author[ikp]{Thomas Krings}
  \author[muenster]{Lukas Lessmann}
  \author[muenster]{Christian Mannweiler}
  \author[ikp]{Dieter Prasuhn}
  \author[ikp]{Steffen Quilitzsch}
  \author[gsi,bochum,ikp]{James Ritman}
  \author[ikp]{Susan Schadmand}
  \author[ikp]{Thomas Sefzick}
  \author[ikp]{Rolf Stassen}
  \author[zea]{Peter Wüstner}

  \cortext[cor]{Corresponding author}

  \address[ikp]{Institut für Kernphysik, Forschungszentrum Jülich, Jülich, 52425, Germany}
  \address[muenster]{Institut für Kernphysik, Universität Münster, Münster, 48149, Germany}
  \address[bochum]{Ruhr-Universität Bochum, Bochum, 44780, Germany}
  \address[gsi]{GSI Helmholtzzentrum für Schwerionenforschung GmbH, Darmstadt, 64291, Germany}
  \address[zea]{Zentralinstitut für Engineering, Elektronik und Analytik, Forschungszentrum Jülich, Jülich, 52425, Germany}

  \begin{abstract}


    The KOALA experiment measures the differential cross section
    of (anti)proton-proton elastic scattering over a wide range of four-momentum
    transfer squared $0.0008 < |t| < \SI{0.1}{\tmom}$.
    The experiment is based on fixed target kinematics and uses an internal hydrogen cluster jet target.
    The wide range of $|t|$ is achieved by measuring the total kinetic energy of the recoil
    protons near \SI{90}{\degree} with a recoil detector, which consists of silicon and
    germanium single-sided strip sensors.
    The energy resolution of the recoil detector is better than
    \SI{30}{\keV} (FWHM).
    A forward detector consisting of two layers of plastic scintillators measures the
    elastically scattered beam particles in the forward direction close to the beam
    axis.
    It helps suppress the large background at small recoil angles and
    improves the identification of elastic scattering events in the low $|t|$ range.
    The KOALA setup has been installed and commissioned at COSY in order to validate the detector by measuring the proton-proton elastic scattering.
    The results from this commissioning are presented here.

  \end{abstract}


\end{frontmatter}


\section{Introduction}
\label{sec:introduction}

A good understanding of the nucleon-nucleon (NN) interaction is one of the principal goals of hadron physics.
Precise and systematic measurements of the differential cross section of the
NN ($\bar{p}p$ or $pp$) elastic scattering provide necessary ingredients
in the modeling of meson production and other nuclear reactions at intermediate energies.
Recent experiments like ANKE \cite{ANKE} and EDDA \cite{EDDA} have filled the gap
in the $pp$ elastic scattering database above \SI{1}{\momentum} in the laboratory reference frame.
However, these experiments only achieved the invariant differential cross section distribution over the region where the nuclear interaction dominates, 
\textit{i.e.}, four-momentum transfer squared $|t| \gtrsim \SI{0.02}{\tmom}$.
Data with smaller \(|t|\), which the Coulomb-Nuclear Interference (CNI) is
dominant, is still missing and is needed to obtain a more accurate estimate of
the total cross section \({\sigma}_{tot}\), the slope parameter \(b\) and the
real to imaginary amplitude ratio \(\rho\) \cite{RevModPhys.57.563}.

\begin{figure*}[htbp]
	\centering
	\includegraphics[width=0.8\textwidth]{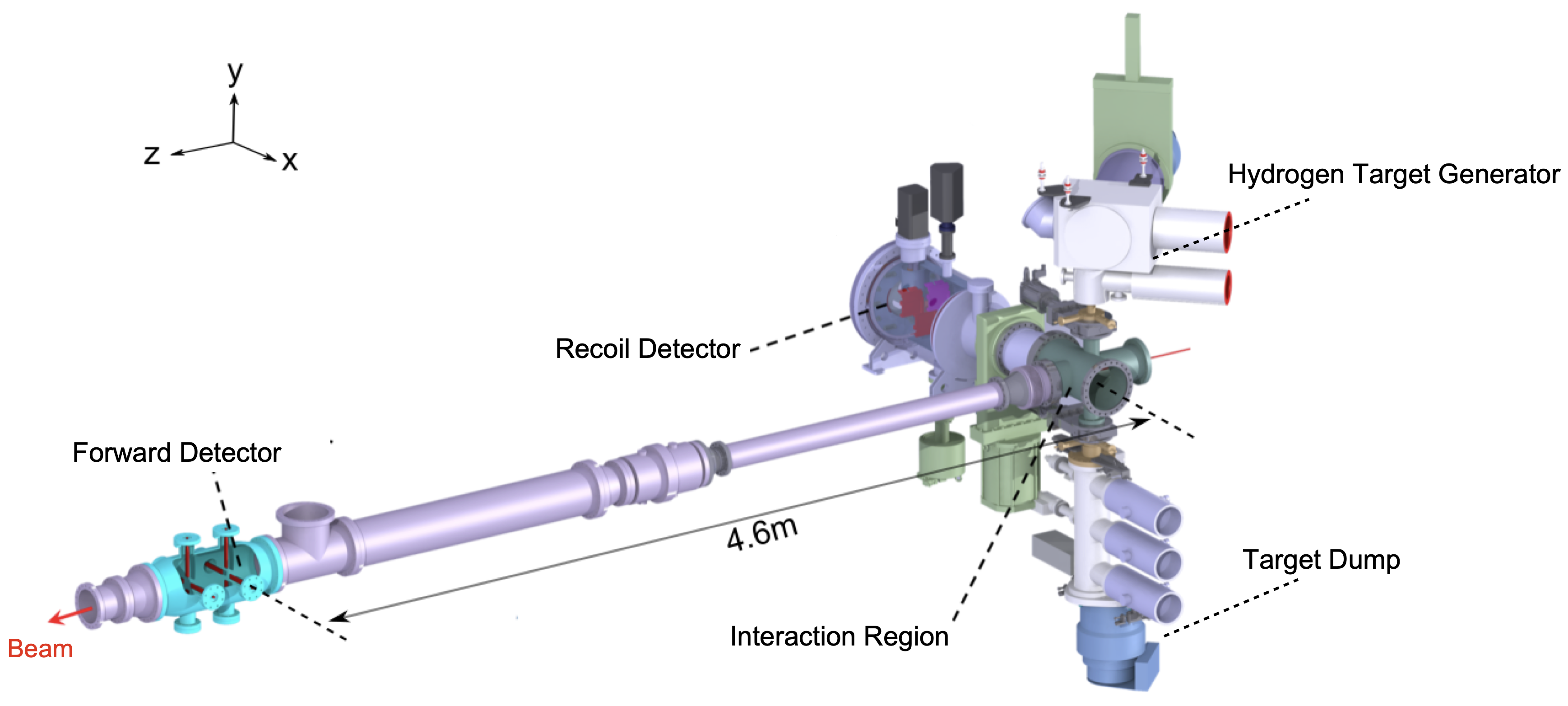}
	\caption{3D visualization of the KOALA setup at COSY.}
	\label{fig:setup}
\end{figure*}

The KOALA experiment is a fixed-target experiment that will measure the
differential cross-section of $\bar{p}p$ or $pp$ elastic scattering
over $0.0008 < |t| < \SI{0.1}{\tmom}$.
Due to the identical kinematics for $\bar{p}p$ and $pp$ elastic scattering, the
same setup can be used in both measurements.
It will measure $\bar{p}p$ elastic
scattering at the future HESR ring of FAIR \cite{FAIR} in the beam momentum range from
\SIrange{1.5}{15}{\momentum}.
KOALA was commissioned with the proton beam from COSY \cite{COSY} to measure $pp$ elastic scattering in the beam momentum range from \SIrange{1.5}{3.4}{\momentum}.
Over this beam momentum range, KOALA covers the kinematic regions where the
Coulomb interaction, the CNI and the nuclear interaction have different relative importance.
The so-called Coulomb normalization method \cite{bernard1987real,jenni2008atlas}
can be used within the Coulomb kinematic region.
This enables the possibility to determine \({\sigma}_{tot}\), \(b\), \(\rho\) as well as
the absolute luminosity by analyzing the characteristic shape of the $dN/dt$
spectrum (i.e. relative differential cross section) \cite{koala_article}.
The absolute differential cross section can be normalized by the data in the
Coulomb region.

Due to limitations from the beam pipe aperture and the beam emittance,
it is extremely difficult to measure the forward emitted elastically scattered particle over a wide range of scattering angles.
The recoil measurement technique, \textit{i.e.}, the precise measurement of both the recoil angle and the kinetic energy of the recoil proton, 
is used to determine the differential elastic scattering cross section in KOALA.
Identification of the elastic scattering events is based on the correlation between the recoil angle and the kinetic energy.
A recoil detector based on this idea has already been built and the method of
recoil measurement technique was verified using proton beam
\cite{koala_article,recoil_article}.
However, it was found in these tests that the identification of elastic events at small recoil angles was limited by the low-energy background events.

The background at low kinetic energy is mainly from inelastic scattering processes.
In order to sufficiently suppress the background across the desired $|t|$ range,
a forward hodoscope based on fast-timing plastic scintillator is
employed to measure the elastically scattered beam particle near the beam axis.
This full setup of KOALA consisting of the recoil detector, the newly-built
forward detector, the cluster jet target and other upgraded components is installed at COSY.
Several tests using proton beams have been carried out to verify the design and the performance.
In the following sections,  the full system of KOALA is described and the preliminary results from the beam commissioning are presented.

\section{Experimental setup at COSY}
\label{sec:setup}

KOALA is installed at a straight section in the COSY storage ring.
The setup consists of three arms as shown in Fig. \ref{fig:setup}: the hydrogen
cluster jet target arm, the recoil arm and the forward arm.
All components in these arms are installed in the same vacuum volume as the
proton beam.

The hydrogen cluster jet target connects the scattering chamber vertically (Y-axis) with the target generator above and the target beam dump below.
The recoil arm is oriented along the X-axis.
The chamber holding the recoil detector is separated from the scattering chamber by a vacuum gate valve.
This valve is used for staged pumping of the recoil chamber during the
preparation of the experiment and separates the detector and the beam vacuum when
either is not within the nominal pressure range for operation.
The forward chamber is located about \SI{4.6}{\meter} downstream from the
central axis of the cluster target jet.
It is connected to the scattering chamber by two beam pipes with diameters of \SI{90}{\mm} and \SI{200}{\mm}.
\subsection{Hydrogen cluster jet target}
\label{sec:target}

A thin hydrogen cluster jet target \cite{cluster_target, cluster_target_new}, which can be operated under ultra high vacuum conditions, is critical for the success of KOALA.
First off, the energy loss of recoil protons should be minimal before impinging
on the recoil sensor so that an accurate determination of its kinetic energy is possible.
Second, the thickness of the cluster jet beam along the COSY beam axis determines
the spread of the vertex distribution of the beam-target interaction.
Since there is no tracking device in KOALA, the spread will deteriorate the
angular resolution as well as the shape of the energy spectrum.

The hydrogen cluster jet target was used previously in the ANKE experiment and has been upgraded for KOALA.
A special collimator has been built to achieve a much
smaller thickness of the cluster jet beam along the beam axis.
The cluster beam profile was measured by using a probing rod when no accelerator beam was circulating inside COSY.
A typical thickness of \SI{1.5}{\mm} (FWHM) at the interaction point (IP) was determined.
The areal density is estimated to be \SI{e14}{\atom\per\cm\squared}, which
enables experiments in COSY with luminosities of $L \approx 10^{30}\si{cm^{-2}s^{-1}}$.

\subsection{Recoil detector}
\label{sec:recoil}

The recoil detector records the recoil proton from the
elastic scattering and measures its total kinetic energy.
For elastic scattering of two particles with the same mass,
the kinetic energy of the recoil proton \(T_p\) is proportional to $|t|$ with \(|t| = 2m_pT_p\), where \(m_p\) is the proton mass.
For $|t|_{max}=\SI{0.1}{\tmom}$, \(T_p \approx \SI{54}{\MeV}\).
Thus, the recoil detector is designed to measure deposited energies of up to \SI{60}{\MeV}.

\begin{figure}[htbp]
  \centering
  \includegraphics[width=0.35\textwidth]{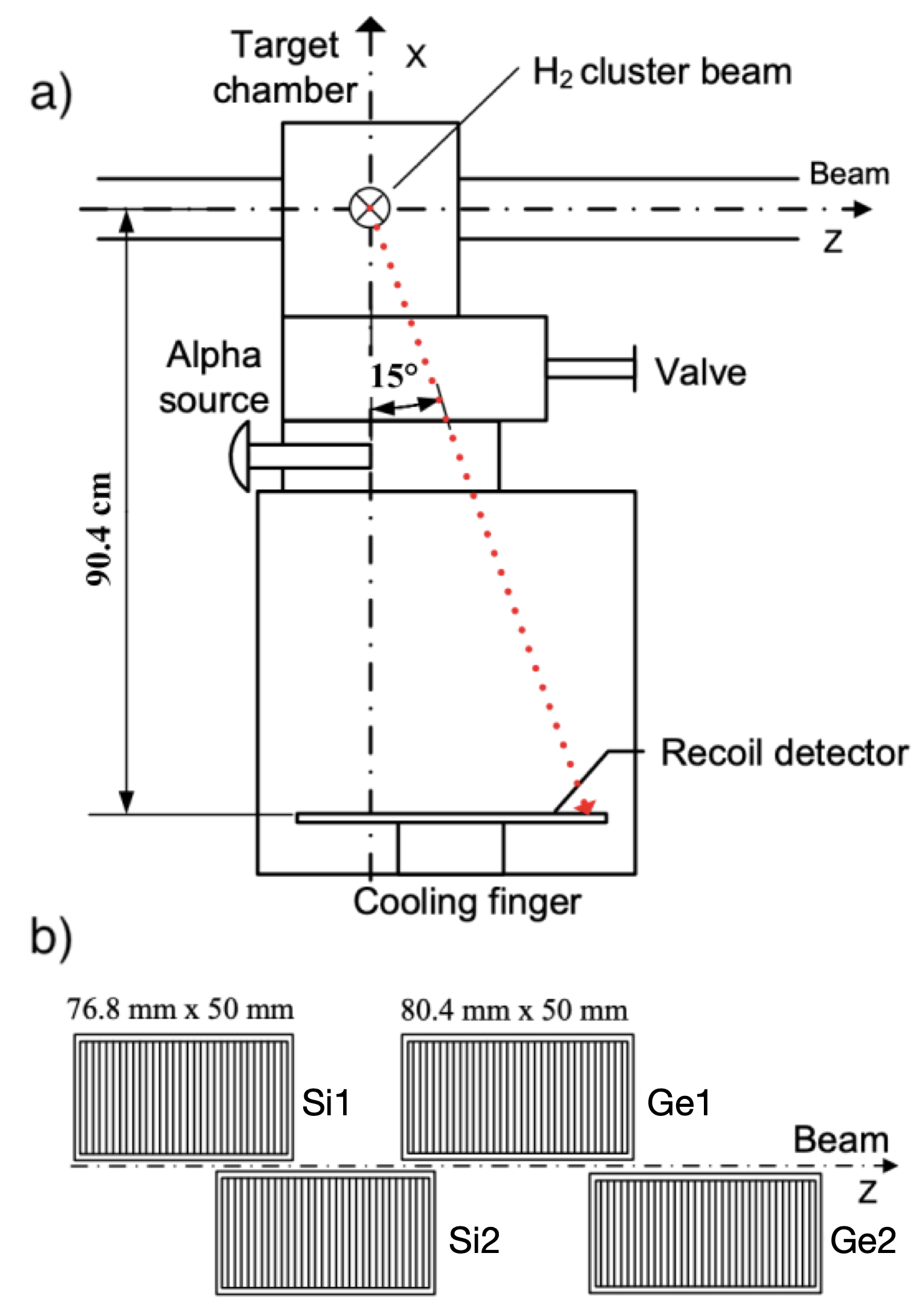}
  \caption{(a) Schematic view of the recoil detector configuration, seen from the
    top; (b) Layout of the four recoil sensors, seen from the interaction point.}
  \label{fig:recoil_schematic}
\end{figure}

The recoil detector consists of two single-sided silicon strip sensors and two
single-sided germanium strip sensors with a layout shown in Fig. \ref{fig:recoil_schematic}.
The detector plane is installed \SI{90.4}{\cm} away from the beam-target center
and covers a range of recoil angle $\alpha$ ($=\SI{90}{\degree}-\theta$) from \SIrange[range-units=repeat]{-1.8}{15}{\degree}.
The sensors are arranged along the beam axis with staggered placement in the Y-axis.
The sensors have different thicknesses: \SI{1}{\mm} for Si1
and Si2, \SI{5}{\mm} for Ge1 and \SI{11}{\mm} for Ge2.
The silicon sensors have a sensitive area of $76.8 \times \SI{50}{\mm\squared}$, which is
segmented into 64 strips of \SI{1.2}{\mm} pitch.
The germanium sensors have a sensitive area of \(80.4 \times \SI{50}{\mm\squared}\), which is segmented into 67 strips of \SI{1.2}{\mm} pitch.
By design, neighboring sensors have an overlapping region, which is symmetric
relative to the beam axis (Si1 and Si2 have 20 overlapping strips, Si2 and Ge1
have 9 overlapping strips, Ge1 and Ge2 have 10 overlapping strips).
The overlapping strips are used to correct the beam position asymmetry.

The recoil sensors are read out by a combination of charge-sensitive preamplifiers (MPR16 for the strips, MPR1 for the rear side) and timing filter amplifiers (MSCF16), from Mesytec \cite{mesytec}. 
MSCF16 integrates a shaping amplifier and a leading edge discriminator in the same module.
Both amplitude and timing signals are extracted from MSCF16 for the energy and time measurement.
Since only 180 readout channels are available, the readout for some of the
strips has been combined, thus there are in effect 48 channels on Si1, 64
channels on Si2, 32 channels each on Ge1 and Ge2, and 4 channels for the rear sides
of the four sensors. 

The germanium sensors need to be operated at low temperature.
Thus, an active cooling system was implemented for the recoil detector.
The temperature of the recoil sensors is monitored by four temperature sensors attached to the detector holder.
The operating temperature can be actively adjusted by a combination of a pulse
tube cold head \cite{pt30} and two heating resistors, which are affixed and thermally coupled to the detector cold plate.
A study of the performance of the recoil sensors using $\alpha$ sources in the
laboratory showed that the working temperature with the best energy resolution is
\SI{125}{\kelvin}.
Under this condition, the energy resolutions of the silicon and germanium
sensors are better than \SI{20}{\keV} and \SI{30}{\keV} (FWHM), respectively.

The $\alpha$ source for the energy calibration (Sec. \ref{sec:calibration}) is also installed inside the recoil chamber and fixed on a linear motion feedthrough rod.
During the experiment, the rod is in a parking position that is hidden from the
recoil detector by the scattering chamber;
when calibration is needed, the rod is pushed to the chamber center and the sources face the recoil sensors directly.
Thus, the recoil detector can be calibrated regularly when no beam is in the storage ring.

More technical information and the detailed performance tests of the recoil detector can be found in \cite{recoil_article}.

\subsection{Forward detector}
\label{sec:fwd}


The forward detector consists of 8 detector modules, which are
grouped into 4 pairs as shown in Fig. \ref{fig:setup}.
They are installed symmetrically along the +X, -X, +Y and -Y axes, with a minimal distance of \SI{3}{\cm} from the beam axis.
The first layer of each pair is installed \SI{4.6}{\meter} downstream of the IP, and the separation between the two layers is \SI{20}{\cm}.
All four pairs are used for beam position tuning and monitoring, but only the
pair on the +X axis is used for the coincidence with the recoil detector.

\begin{figure}[htb]
  \centering
  \includegraphics[width=0.45\textwidth]{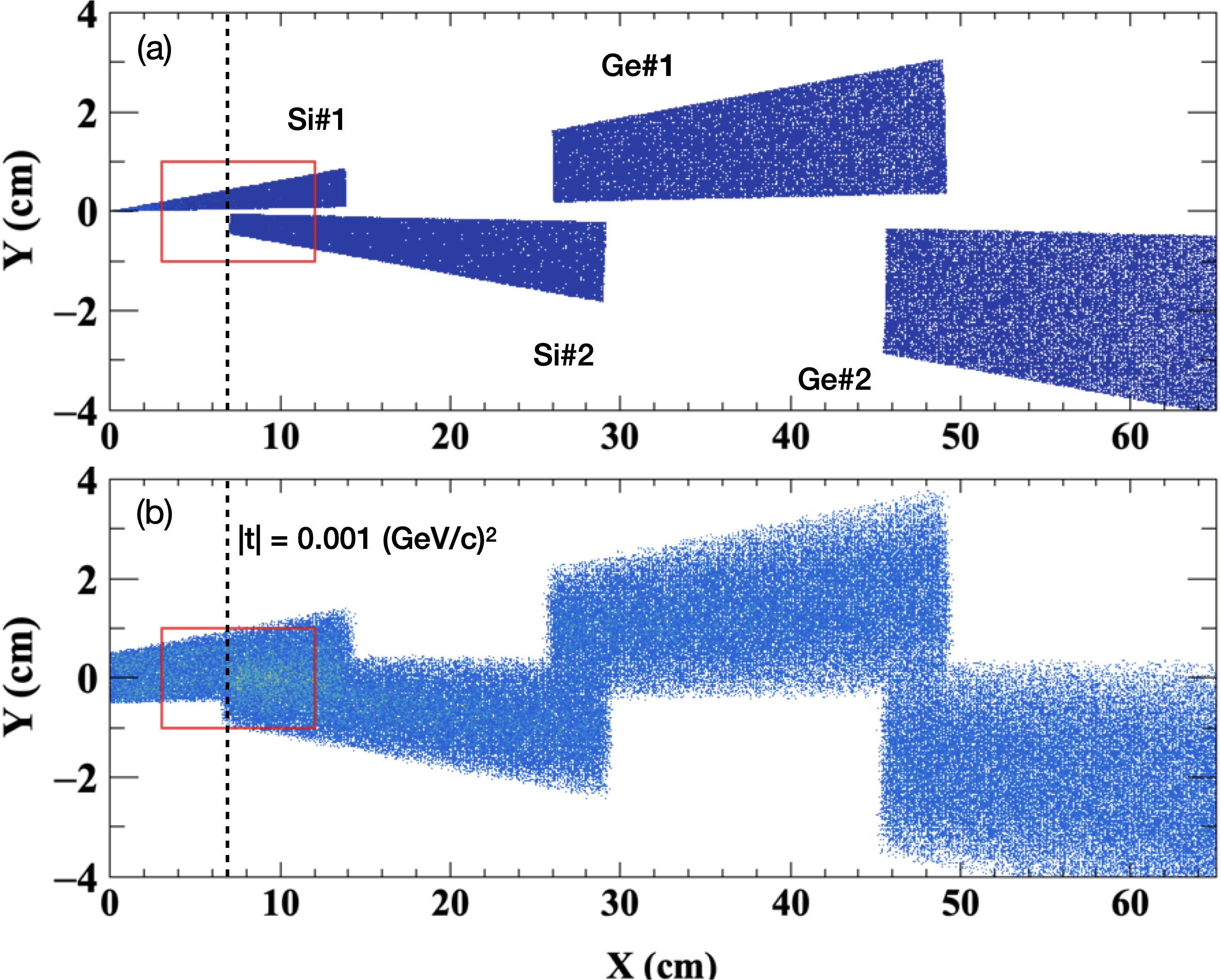}
  \caption{
    Distribution of hit position of the forward scattered particle on a plane at
    $z=\SI{4.6}{\m}$ that is coincident with the recoil particle hitting one of
    the four recoil sensors.
    The red square indicates the fiducial volume of the forward scintillators.
    The vertical dashed line indicates the hit position when $|t| = \SI{0.001}{\tmom}$.
    The results are from the simulation of \SI{2.2}{\momentum}
    $pp$ elastic scattering: a) ideal point-like beam; b) $\SI{10}{\mm}\times\SI{10}{\mm}$ box beam profile.
    }
  \label{fig:forward_acceptance}
\end{figure}

Each detector module is made of a plastic scintillator BC-408 \cite{bc408} with the dimensions $90 (length) \times 20 (width) \times 6
(thickness)\,\si{\mm\tothe{3}}$.
This corresponds to an angular acceptance of the scattering angle $\theta$ from
\SIrange[range-units=repeat]{0.37}{1.24}{\degree}, the upper limit of which is
constrained by the diameter (\SI{20}{\cm}) of the beam pipe.
Due to the kinematics of elastic scattering, coincidences between the forward
detector and the recoil detector are limited not only by the size of the forward
detector but also by the finite emittance of the beam at the IP.
Fig. \ref{fig:forward_acceptance} shows an example of the impact of beam emittance on the
effective momentum transfer coverage.
The fully-covered region shrinks with increasing beam emittance. 
Based on the simulation study, the current width of the forward detector
scintillator only guarantees coincidences between the forward detector and the
recoil detector for $|t| < \SI{0.001}{\tmom}$ at the COSY beam momentum range if the beam spot has a diameter smaller than \SI{7}{\mm}.

\begin{figure}[htbp]
  \centering
  \includegraphics[width=0.32\textwidth]{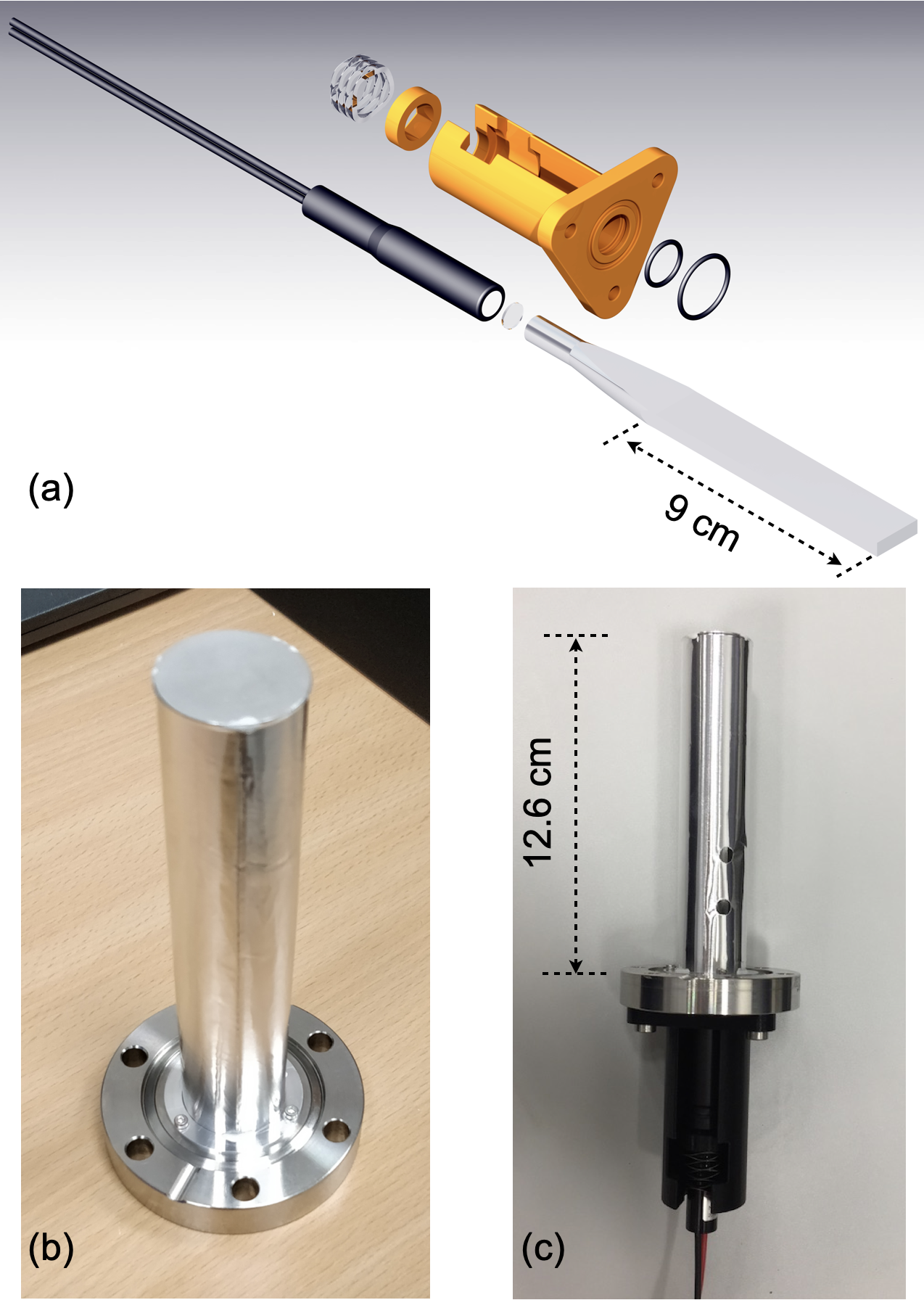}
  \caption{(a) The CAD model of the assembly of a forward detector module; (b) The
    aluminum tube is fixed on the flange inside the vacuum volume; (c) One forward detector module after complete assembly.}
  \label{fig:forward_module}
\end{figure}

The scintillator is read out by the combination of a tapered light guide, a piece of silicone pad and a
photomultiplier tube (Hamamatsu H6900 \cite{hamamatsu}), as shown in Fig. \ref{fig:forward_module} (a).
Each detector module is integrated into a forward chamber flange for usage
in the ultra-high vacuum.
To meet the vacuum requirements inside the beam pipe, the following designs are implemented:
\begin{enumerate}
\item only the light guide and the scintillator are installed inside the forward chamber, the light guide is glued on the open port of the flange as a feedthrough;
\item no wrapping and painting material on the surface of the scintillator, the
  surface is polished to increase the light collection efficiency;
\item a thin aluminum tube of \SI{100}{\micro\meter} thickness is used as
  the light shield to prevent light cross-talk between the scintillators, the tube is screwed on the flange, see Fig. \ref{fig:forward_module} (b);
\item two small holes, which are aligned to the thin side of the scintillator, are opened on the wall of the aluminum tube to speed up vacuum pumping, see Fig. \ref{fig:forward_module} (c).
\end{enumerate}
Other components such as the silicone pad, the PMT and the PMT base are
installed inside a light-tight case outside of the vacuum volume as shown in Fig. \ref{fig:forward_module} (c).

Due to the large gain of the PMT, no front-end electronics are needed for the readout of the forward detector.
The output signal from H6900 is split into two branches: one fed to a
constant fraction discriminator for the time information extraction; the other
for the charge measurement.
The modules were tested in the laboratory using cosmic rays before installation.
In each test, two modules were arranged in a cross layout so that only vertical
cosmic rays were used.
The minimum ionizing particle (MIP) peak is well separated from the pedestal
noise, with the signal to noise ratio (SNR) larger than 50.
The timing resolution of a single module ($\sigma_{module}$) can be deduced from
the time difference distribution of the two modules being tested by $\sigma_{module}=\sigma_{diff}/\sqrt{2}$.
Typical timing resolutions (FWHM) are about \SI{360}{\pico\second}.

\section{Data acquisition system}
\label{sec:daq}

The data acquisition system of KOALA is a VME-based system with multiple types
of digitization modules produced by Mesytec \cite{mesytec}.
For the recoil detector, the amplitude signal from MSCF16 is digitized by a
peak-sensing ADC module MADC-32.
MADC-32 has a 13-bit dynamic range with \SI{12.5}{\micro\second} conversion time.
For the forward detector, the pulses from PMTs are directly fed into a QDC
module MQDC-32 for the charge measurement.
MQDC-32 has a dynamic range of \SI{500}{\pico\coulomb} with 12-bit resolution
and \SI{250}{\nano\second} conversion time.
The timing information from both the recoil and forward detectors are recorded
by a TDC module MTDC-32 with \SI{32}{\pico\second} resolution.
MTDC-32 modules are configured to work in the Start-Stop mode, with the DAQ
trigger signal serving as the common start.
A multi-channel scalar SIS3820 \cite{sis} is also integrated to record
the following key count rates: 1) count rates of the four pairs of the forward detector for beam position monitoring; 2) count rates of the overlapping regions of the recoil detector for asymmetry correction; 3) count rates of the input trigger for DAQ efficiency correction.

\begin{figure*}[tb!]
  \centering
  \includegraphics[width=0.75\textwidth]{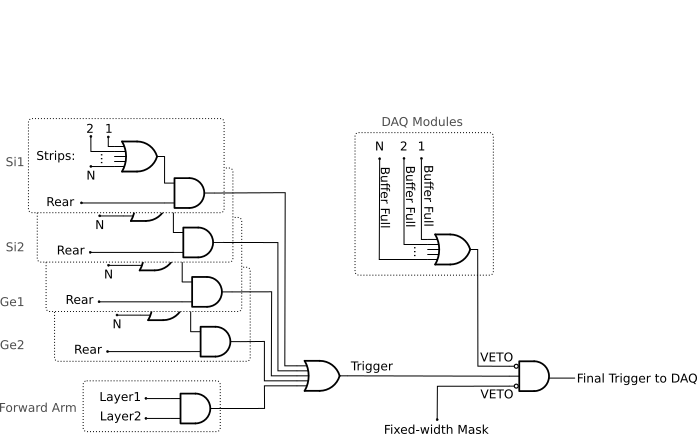}
  \caption{Schematic overview of the KOALA DAQ trigger logic.}
  \label{fig:trigger_logic}
\end{figure*}

KOALA adopts a self-triggering scheme for the trigger logic design.
Each sensor of the recoil detector and +X arm of the forward detector works independently and generates its own trigger. 
The final trigger to the DAQ is a common OR of all sub-detectors, as shown in Fig. \ref{fig:trigger_logic}.
The trigger from the recoil detector sensor is generated by a coincidence between the front-side strips and the rear-side plane, 
and the trigger from the forward detector arm is generated by a coincidence between the two modules in the pair.
Both elastic and inelastic scattering events are recorded in this trigger
design, and the coincidence between the recoil sensor and the forward detector
is carried out in the offline analysis.

Fast readout of the recorded event is crucial for a self-triggered DAQ system.
The asynchronous readout mechanism combined with VME CBLT block read mode is adopted to increase the data throughput.
The digitization modules used in KOALA have an event buffer with a size
larger than \SI{32}{\kibi\byte}.
Digitized events are stored in this buffer before readout, so that the module is
available to digitize the next event without further dead time.
Events will not be readout until the buffer is nearly full.
On the other hand, since the cross section is much higher at smaller recoil
angles, the module buffer for these channels saturates faster than others.
These modules may miss new events while others might not, generating a systematic bias.
To overcome this issue, the buffer-full signal from each module is added to the
trigger logic as a \texttt{VETO} as shown in Fig. \ref{fig:trigger_logic}.

The need for event synchronization arises naturally when using the asynchronous
readout mode.
Timestamp-based synchronization is used to solve this problem.
The modules in the system have a 30-bit counter for recording the timestamp from
the clock signal distributed by a central source.
The source could be either the VME built-in clock (\SI{16}{\MHz}) or an external clock
(lower than \SI{75}{\MHz}).
Currently, the built-in clock of VME backplane bus is used. 
Based on this timestamp, event synchronization is achieved offline.
The other option is adding a fixed-width mask gate into the trigger logic as \texttt{VETO}, see Fig. \ref{fig:trigger_logic}.
The width of the mask gate should be larger than the maximum dead time of all modules.
In this way, the events are effectively synchronized sequentially. 
However, this method may reduce the DAQ efficiency significantly in a high hit-rate environment.

\begin{figure}[htbp]
  \centering
  \includegraphics[width=0.45\textwidth]{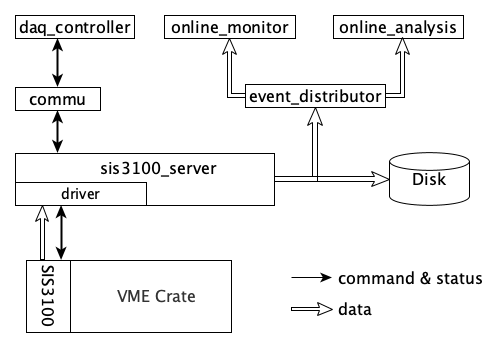}
  \caption{Schematic overview of the design of KoalaEms.}
  \label{fig:koalaems}
\end{figure}

A dedicated DAQ software called KoalaEms is also developed for KOALA.
KoalaEms is a fork of the EMS software \cite{ems}, which is a highly flexible DAQ software framework developed for various experiments at COSY.
Support for the VME controller (SIS3100 \cite{sis}) is integrated and a new
component of the online monitoring is added.
The architecture of KoalaEms is shown in Fig. \ref{fig:koalaems}.
The interface to the DAQ is implemented as \texttt{sis3100\textunderscore server}, the host of which
connects to SIS3100 by an optical link.
The control and status information from/to the GUI \texttt{daq\textunderscore controller} is mediated by a component called \texttt{commu}.
The data flow from VME crate is split into two branches:
\begin{enumerate*}[label=(\roman*)]
\item \texttt{data\textunderscore out\textunderscore di\allowbreak sk}: save the raw data onto disk;
\item \texttt{data\textunderscore out\textunderscore stream}: stream out to \texttt{event\textunderscore distributor}.
\end{enumerate*}
The data stream is forwarded by \texttt{event\textunderscore distributor} to various consumption hosts for usages like online monitoring and online analysis.
Both \texttt{commu} and \texttt{event\textunderscore distributor} support socket connection and \texttt{event\textunderscore distributor} also supports multiplexing streaming.
All components shown in Fig. \ref{fig:koalaems} can be hosted in different PCs
and a new consumption host to the data stream can be integrated whenever needed.

\section{Software framework}
\label{sec:software}

\begin{figure}[htbp]
  \centering
  \includegraphics[width=0.48\textwidth]{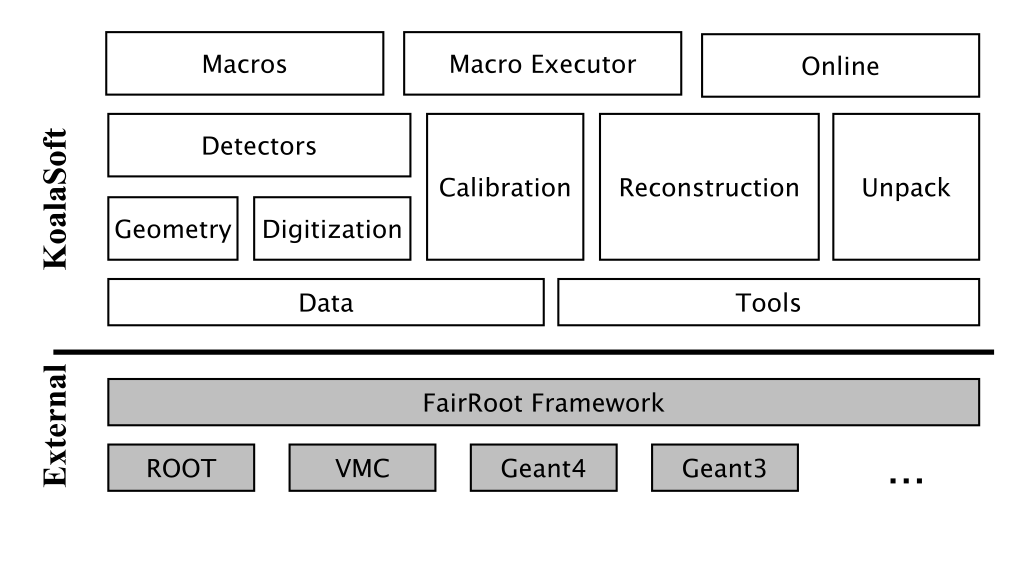}
  \caption{Overview of the KoalaSoft components.}
  \label{fig:koalasoft}
\end{figure}

A software framework called KoalaSoft has been developed for the simulation, calibration, reconstruction and analysis jobs of the KOALA experiment.
It is built upon the FairRoot \cite{fairroot} framework, which implements a
simulation environment based on the VMC \cite{vmc} library and an analysis
environment based on ROOT's task concept \cite{root}.
The components stack of KoalaSoft is shown in Fig. \ref{fig:koalasoft}.

By using VMC it is possible to choose between Geant3 or Geant4 as the simulation engine without changing other components in KoalaSoft.
Geometry models of the recoil detector and the forward detector are implemented
using the TGeo library in ROOT.
Jobs such as digitization, calibration and reconstruction are divided into multiple smaller steps, each of which is represented by a single task.
Tasks are chained together later in a ROOT macro to compose a full job. 
ROOT macros are the interface for the end user using KoalaSoft.
Macros for common jobs are pre-configured and distributed along with KoalaSoft.
Users can also compose their own specific jobs for analysis.
Additionally, a binary macro executor is provided to run jobs directly from the command line. This may be useful in batch processing.

The same chain of tasks can be used for the analysis of both the simulated data
and the raw data from the DAQ.
This is accomplished by the \texttt{Unpack} component, which can decode and transform the raw binary data into the same format as the output from simulation jobs.
Thus, algorithms developed, tested and verified using simulated data can be applied to experimental data seamlessly.
This saves much effort in the development and maintenance of algorithms.
Both the offline disk data and the online streaming data are correctly handled by \texttt{Unpack} and an online monitoring program is developed based on it.

\section{Reconstruction}
\label{sec:reconstruction}

\subsection{Time walk correction}
\label{sec:timewalk}

\begin{figure}[htbp]
  \centering
  \includegraphics[width=0.45\textwidth]{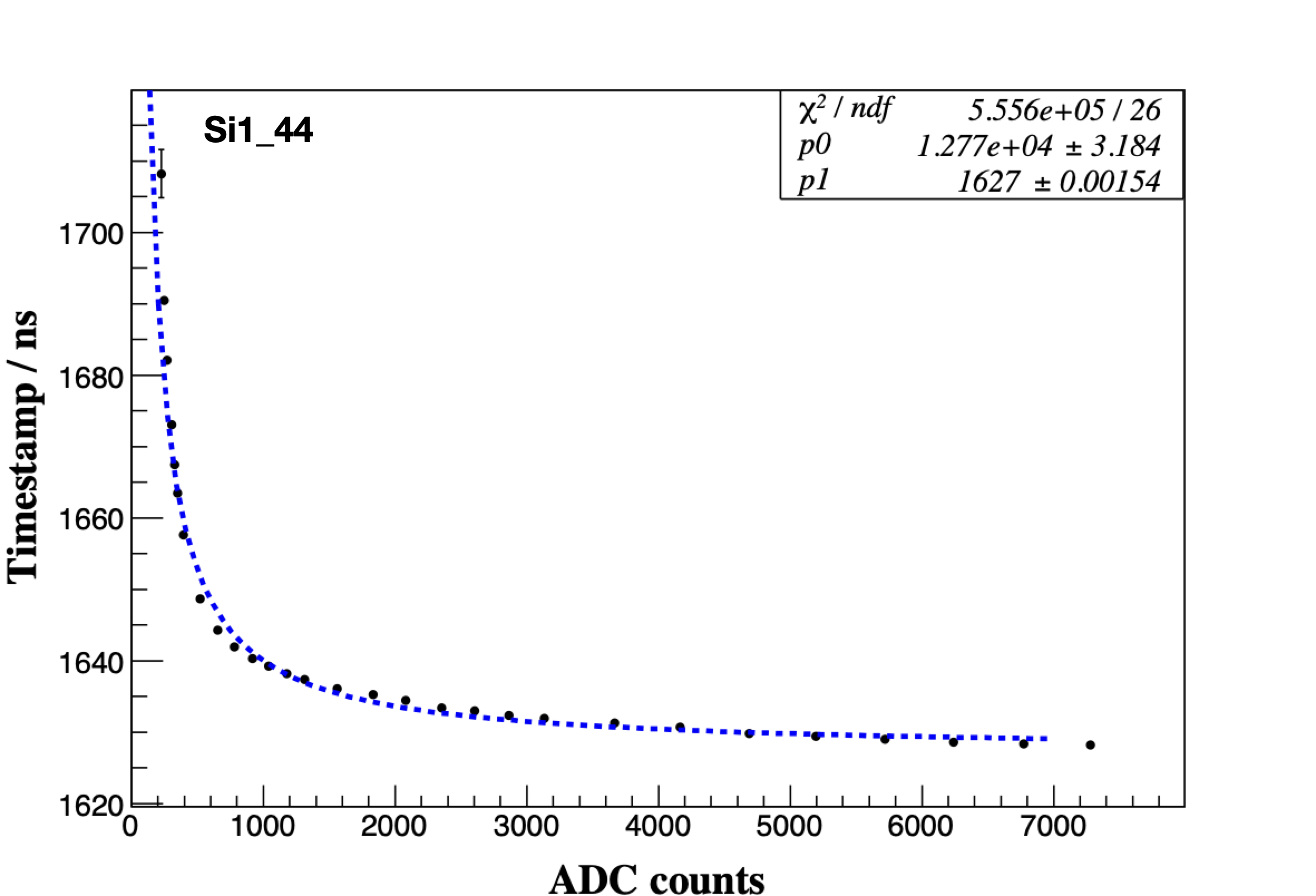}
  \caption{Typical result from the time-walk calilbration, plotted for a single
    strip on the Si1 sensor.}
  \label{fig:timewalk}
\end{figure}

Time walk of the leading edge discriminator used in the recoil detector needs to be corrected to optimize the timing resolution.
The calibration of the time-walk effect is carried out with a precision pulse generator (ORTEC Model 419 \cite{ortec}). 
The output of the pulser is split into two branches: one is fed into a constant fraction discriminator to generate the reference time;
the other is connected to the detector channel that is being calibrated. 
By scanning the pulser over a wide range of amplitudes, the time-walk effect is measured.
An example is shown in Fig. \ref{fig:timewalk}, where the result has been fit with \(y=p_0 x^{-1} + p_1\). 
The time walk is corrected by subtracting \(\Delta t = p_0/ADC\) from the measured timing value.

The difference of the fit parameters \(p_1\) indicates the delay time
difference between different channels, which is caused by the signal routing length variation.
These offsets are also corrected in the reconstruction to align the timing from different channels.

\subsection{Energy calibration}
\label{sec:calibration}

\(^{239}Pu\), \(^{244}Cm\), \(^{241}Am\), with main $\alpha$ decay energies
of \SIlist{5156.59;5804.83;5485.56}{\keV}, respectively \cite{nuclear_data}, are
used for the energy calibration of the recoil detector.
Decays with smaller branching ratios are also utilized if they are well separated from the main peaks.

Two aspects need special consideration in the calibration.
First off, the recoil sensors have a thin protective layer on the surface and this
causes energy loss before particles enter the fiducial volume of the sensor.
The thickness of the protection layer has been determined in the laboratory
\cite{recoil_article} and the energy loss of $\alpha$ particles can be estimated
with the LISE++ program \cite{LISE}.
The estimated energy losses of $\alpha$ particles incident at a normal angle to
the detector surface are listed in Tab. \ref{tab:dead_layer} for each sensor.
The effective energy deposit in the fiducial volume is then corrected based on the recoil angle, where each strip is located.
\begin{table}[h!]
  \centering
  \caption{Energy loss (\si{\keV}) in the protection layer for an alpha particle
    at nominal incidence angle.}
  \label{tab:dead_layer}
  \begin{tabular}{cccccc}
    \hline
    \(E_{\alpha} (keV)\) & \(\Delta E_{Si1}\) & \(\Delta E_{Si2}\) & \(\Delta E_{Ge1}\) & \(\Delta E_{Ge2}\) \\
    \hline
    5156.59 & 11.51 & 11.51 & 110.00 & 111.00 \\
    5485.56 & 11.01 & 11.01 & 105.00 & 106.00 \\
    5804.83 & 10.52 & 10.52 & 99.00  & 100.00 \\
    \hline
  \end{tabular}
\end{table}

\begin{figure}[bht!]
  \centering
  \includegraphics[width=0.45\textwidth]{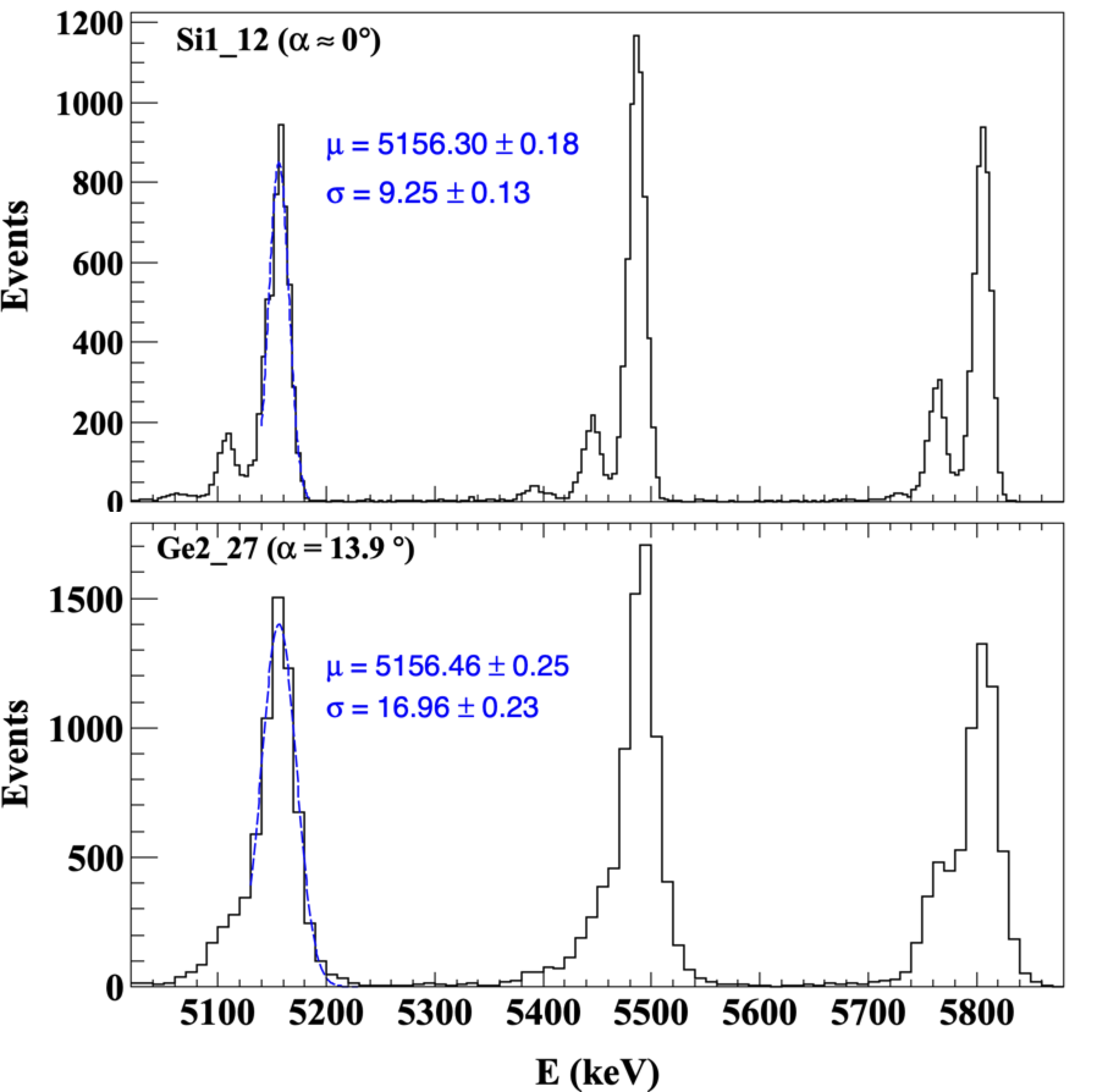}
  \caption{Energy spectra of \(\alpha\) sources for two channels at small (up)
    and large (down) recoil angles. The main peak from $^{239}Pu$ at
    \SI{5156.59}{\keV} in each spectrum has been fit with a Gaussian.}
  \label{fig:alpha_spectrum}
\end{figure}

Secondly, the gain setting of each readout channel is optimized for the energy range covered by the corresponding strip.
The gain difference varies by up to a factor $\mathtt{\sim}$10.
Thus, the resolution is worse at large recoil angles than at small angles.
The recorded $\alpha$ energy spectra by one channel close to the IP and one
channel close to the far edge of the recoil detector are shown in Fig. \ref{fig:alpha_spectrum}.
The minor decay modes can not be recognized with a small gain setting (the bottom frame
in Fig. \ref{fig:alpha_spectrum}), while they are clearly seen with a large gain
setting (the upper frame in Fig. \ref{fig:alpha_spectrum}).
The worse resolution brings larger uncertainty in the calibration, since only three energy points can be used.

\begin{figure}[b!]
  \centering
  \includegraphics[width=0.45\textwidth]{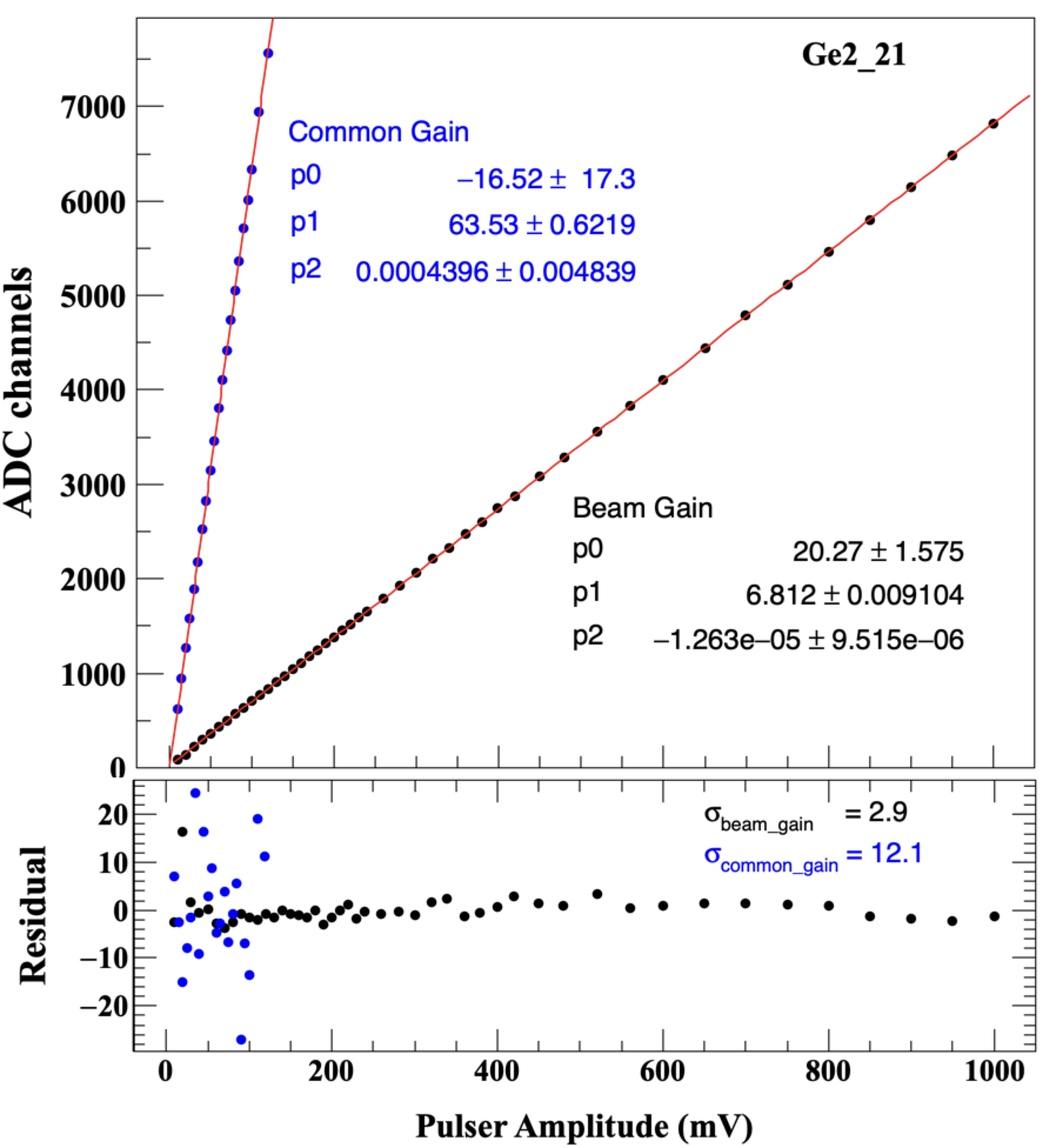}
  \caption{The results of a quadratic fit to the linearity curves of a recoil
    detector channel. The bottom frame shows the residuals between the data
    points and the linear part of the quadratic function.}
  \label{fig:electronic_linearity}
\end{figure}

To minimize this uncertainty, a common gain setting, which is optimized for the separation of the \(\alpha\) peaks, is set for all channels.
The calibration is carried out indirectly as follows:
\begin{enumerate}
\item \(\alpha\) source calibration is carried out under the common gain setting;
\item the linearity curves at both the common gain and the so-called beam gain
  setting used in the experiment are calibrated using the pulse generator ORTEC Model 419;
\item the $\alpha$ peak at the beam gain setting is then deduced
  from the linearity curve at a pulser amplitude which generates the same peak
  value as the $\alpha$ source when using the common gain setting;
\item the peak responses are fit linearly against the corresponding $\alpha$ energies
  to get the calibration parameters.
\end{enumerate}
The readout electronics of the recoil detector have a good linearity in both common gain and beam gain settings.
Fig. \ref{fig:electronic_linearity} shows the quadratic fit results to the linearity curves for a readout channel of Ge2.
In both cases, the quadratic terms are negligibly small and the residuals
between the measurement data points and the linear part have small dispersions
($\sigma=\SI{2.9}{channels}$ for the beam gain setting, $\sigma=\SI{12.5}{channels}$ for the common gain setting).
Thus, linear fitting is adopted in all the calibration steps.
The fitting parameters from the last step are used to convert ADC channel into
energy value in the reconstruction.

\subsection{Clustering}
\label{sec:clustering}

Clustering is necessary to correctly reconstruct the deposited energy:
\begin{enumerate*}[label=(\roman*)]
\item charge sharing is an intrinsic characteristic of the solid-state strip detector, especially for the tracks that hit the region close to the border between adjacent strips;
\item tracks with large recoil angles may penetrate several strips before stopping.
\end{enumerate*}
The results from the simulation and the beam commissioning both show that more
than \SI{50}{\percent} of the tracks that hit the Ge2 sensor have a hit multiplicity
larger than 1.
The hit multiplicity distributions in the four
recoil sensors are listed in Tab. \ref{tab:multiplicity}.
\begin{table}[htb!]
  \centering
  \caption{Hit multiplicity distribution for the recoil sensors recorded for
    proton beam at $P_{beam} = \SI{2.2}{\momentum}$.}
  \label{tab:multiplicity}
  \begin{tabular}{@{} *{5}{c} @{}}
    \hline
    \headercell{\\Sensor} & \multicolumn{4}{c@{}}{Mutliplicity (\si{\percent})} \\
    \cmidrule(l){2-5} & 1 & 2 & 3 &  4 \\
    \midrule
    Si1 & 95.51 & 4.38 & 0.08 & 0.02 \\
    Si2 & 92.46 & 7.21 & 0.24 & 0.06 \\
    Ge1 & 81.06 & 16.61 & 1.51  & 0.44 \\
    Ge2 & 45.41 & 46.30 & 3.98  & 1.67 \\
    \hline
  \end{tabular}
\end{table}

The clustering algorithm is designed to minimize the noise contribution while
not introducing a selection bias between the low energy and high energy elastic events.
The steps of clustering in KOALA are as follows:
\begin{enumerate}
\item Determine the noise level $\sigma_{noise}$. An event sample which is
  triggered only by the forward detector is selected.
  The pedestal of the energy spectrum is fit with a Gaussian to extract $\sigma_{noise}$ for each
  readout channel.
  \begin{figure}[b]
    \centering
    \includegraphics[width=0.45\textwidth]{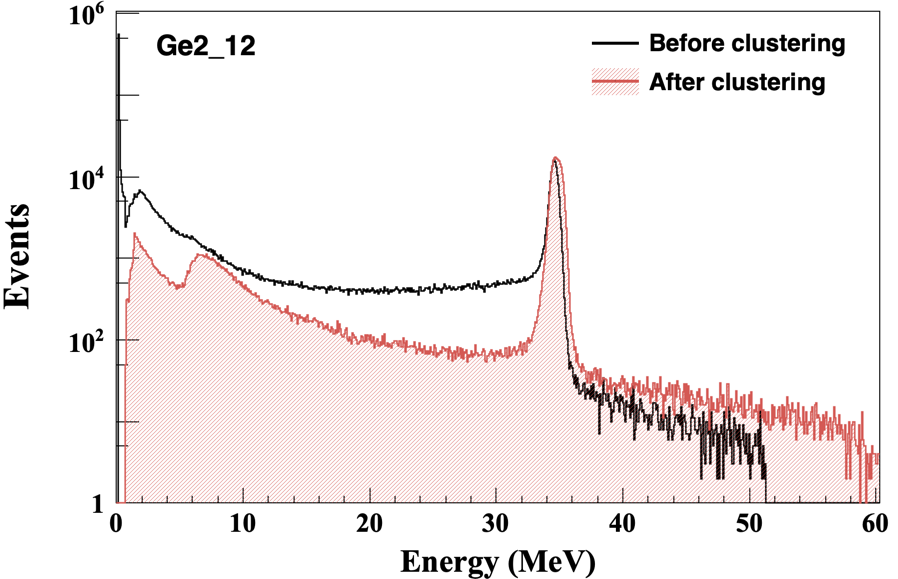}
    \caption{The recoil energy spectrum measured by a channel of the Ge2 sensor before (black) and after (red) the clustering.}
    \label{fig:clustering}
  \end{figure}
\item Delete noise hits with energy below a low threshold of $2\sigma_{noise}$.
\item Combine adjacent hits that are above threshold into clusters.
\item Delete noise clusters. The noise level of a cluster with $n$ hits is evaluated as
  $\sigma_{cluster} = \sqrt{\sum_{i=1}^n{(\sigma_{noise}^i)^2}}$, where $\sigma_{noise}^i$ is
  the noise level of the $i_{th}$ strip in the cluster. Clusters with energy lower than $5\sigma_{cluster}$ are considered to be noise-induced and deleted.
\item Delete clusters without a seed hit. The seed hit is defined as the
  hit whose energy is above the trigger threshold of the corresponding readout
  channel. Since a low trigger threshold is used ($\sim\SI{50}{\keV}$), valid clusters need to have at least one seed hit.
\end{enumerate}
The hit position and hit time of the cluster are both extracted from the first
hit in the cluster to keep the most accurate information about the recoil angle.

A comparison between the energy spectrum before and after clustering is shown in Fig. \ref{fig:clustering}.
After clustering, a large fraction of the hits in the plateau below the elastic peak, which exists in the spectrum before clustering, correctly migrates into the elastic peak.
The charasteristic shape of the background spectrum also shows up more explicitly.
Very noticible is the appearance of a peak at about \SI{7}{\MeV} in this figure,
which is caused by MIPs (mostly \Pgppm) from inelastic events.
A correct description of the background shape (see Sec.
\ref{sec:recoil_performance}) is only possible after clustering.

\subsection{Alignment of the recoil sensors}
\label{sec:alignment}

During installation, the position of the KOALA setup was adjusted to a precision
of \SI{0.1}{\mm} relative to the COSY coordinate system by using a laser positioning system.
However, the relative misalignment between the recoil sensors can't be
determined in this way because they are mounted together on a common sensor holder.
On the other hand, the exact location of the IP needs to be determined for each experiment since
it may vary with the operation conditions of the cluster jet target.
Thus, an alignment based on the experiment data is needed to accurately deduce the recoil angle of the elastic scattering events.

\begin{figure}[th!]
  \centering
  \includegraphics[width=0.45\textwidth]{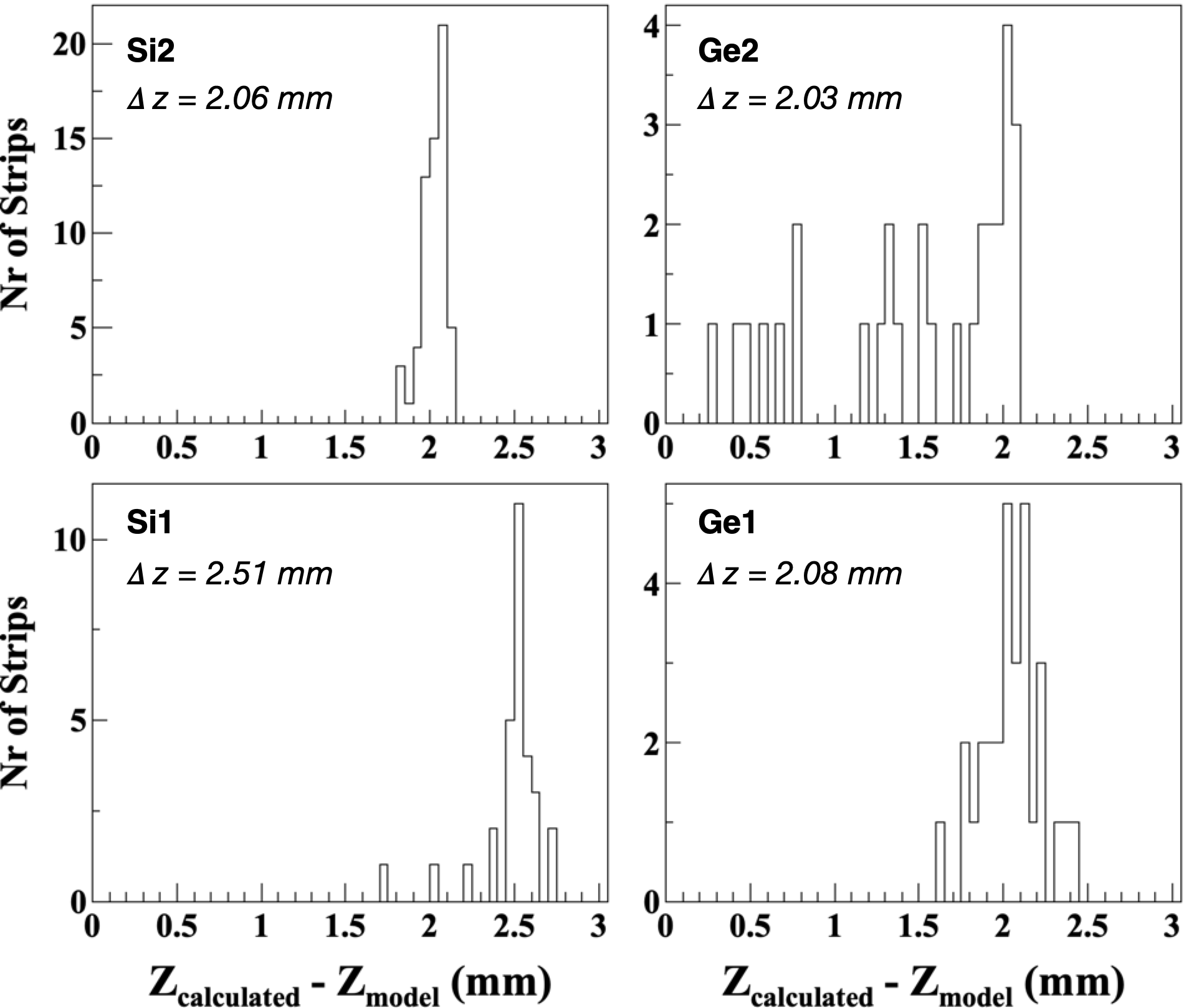}
  \caption{Histograms of the difference between the calculated position from the
    elastic peak energy and the nominal position from the geometry model.
    The obtained alignment parameters are
    \SIlist[list-units=single]{2.51;2.06;2.08;2.03}{\mm} for Si1, Si2, Ge1, Ge2, respectively.
    This result is from the data obtained at $P_{beam} =
    \SI{2.2}{\momentum}$.}
  \label{fig:alignment}
\end{figure}

Due to the layout of the recoil sensors, only the alignment along
the beam axis, \textit{i.e.}, Z-axis in the COSY coordinate system, is necessary.
The alignment parameter for each sensor is determined as follows:
\begin{enumerate}
\item the peak energy of elastic scattering events on each strip is extracted
  (see Sec. \ref{sec:recoil_performance} and Sec. \ref{sec:tofe_selection}) and converted to a calculated position along the beam axis based on the elastic
  scattering kinematics;
\item for strips from the same sensor, the difference between the calculated
  position and the nominal position of each strip from the ideal geometry model is filled into the same histogram, as shown in Fig. \ref{fig:alignment};
\item the peak region in the histogram is fit with a Gaussian to get the offset
  of each sensor compared to the nominal position.
\end{enumerate}
These offset values are the corresponding alignment parameters that are later applied in the reconstruction.
Thus, the recoil sensors are aligned so that the IP is at the origin of the
laboratory coordinate system and the misalignment between the sensors is
simultaneously corrected.

\section{Commissioning with proton beam}
\label{sec:result}

Tests were carried out using the proton beam at COSY with momenta of
\SIlist[list-units=single]{2.2;2.4;2.6;3.0}{\momentum}.
In the following, the results from \SI{2.2}{\momentum} are shown as an example.

\subsection{Beam condition}
\label{sec:beam}
\begin{figure}[h]
  \centering
  \includegraphics[width=0.45\textwidth]{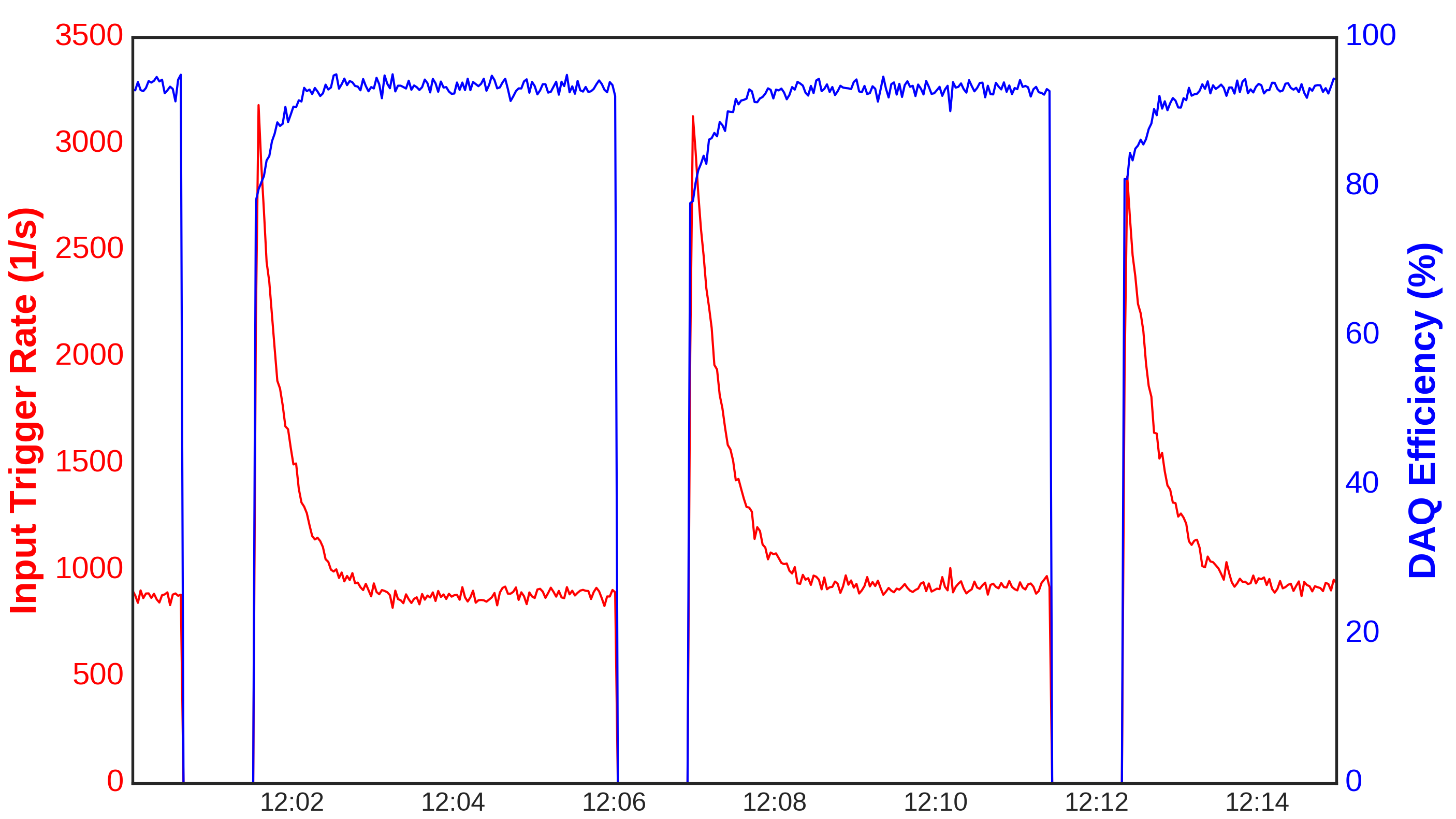}
  \caption{Time variation of the total input trigger rate (red) and the ratio between the accepted and total input trigger rate (blue), which is a direct measure of the DAQ efficiency.}
  \label{fig:beam}
\end{figure}

Typically there are $10^{10}$ protons stored in COSY, which has a circumference of
\SI{184}{\meter}.
The storage time for a beam cycle is about \SI{300}{\second}, as shown in Fig. \ref{fig:beam}.
To minimize the beam emittance, stochastic cooling \cite{cooling} was applied in these tests.

The experiment is only carried out during the storage cycle, which is achieved by two
gates generated by the beam cycle.
One gate is used to protect the PMTs of the forward detector during injection.
It ramps down the high voltage supply \SI{5}{\second} before the end of the
storage cycle and ramps up the high voltage supply \SI{5}{\second} after the start of the storage cycle.
The other gate has a narrower width and is used to control the DAQ system so
that data is not recorded during injection and beam dump.

The DAQ efficiency is seen in Fig. \ref{fig:beam} to strongly depend upon the
instantaneous trigger rate.
The DAQ was operated under the mask-gate based event synchronization.
A maximum efficiency of \SI{93}{\percent} is reached at a trigger rate of \SI{850}{\event\per\second}.

\subsection{Performance of the forward detector}
\label{sec:fwd_performance}

\begin{figure}[b!]
  \centering
  \includegraphics[width=0.48\textwidth]{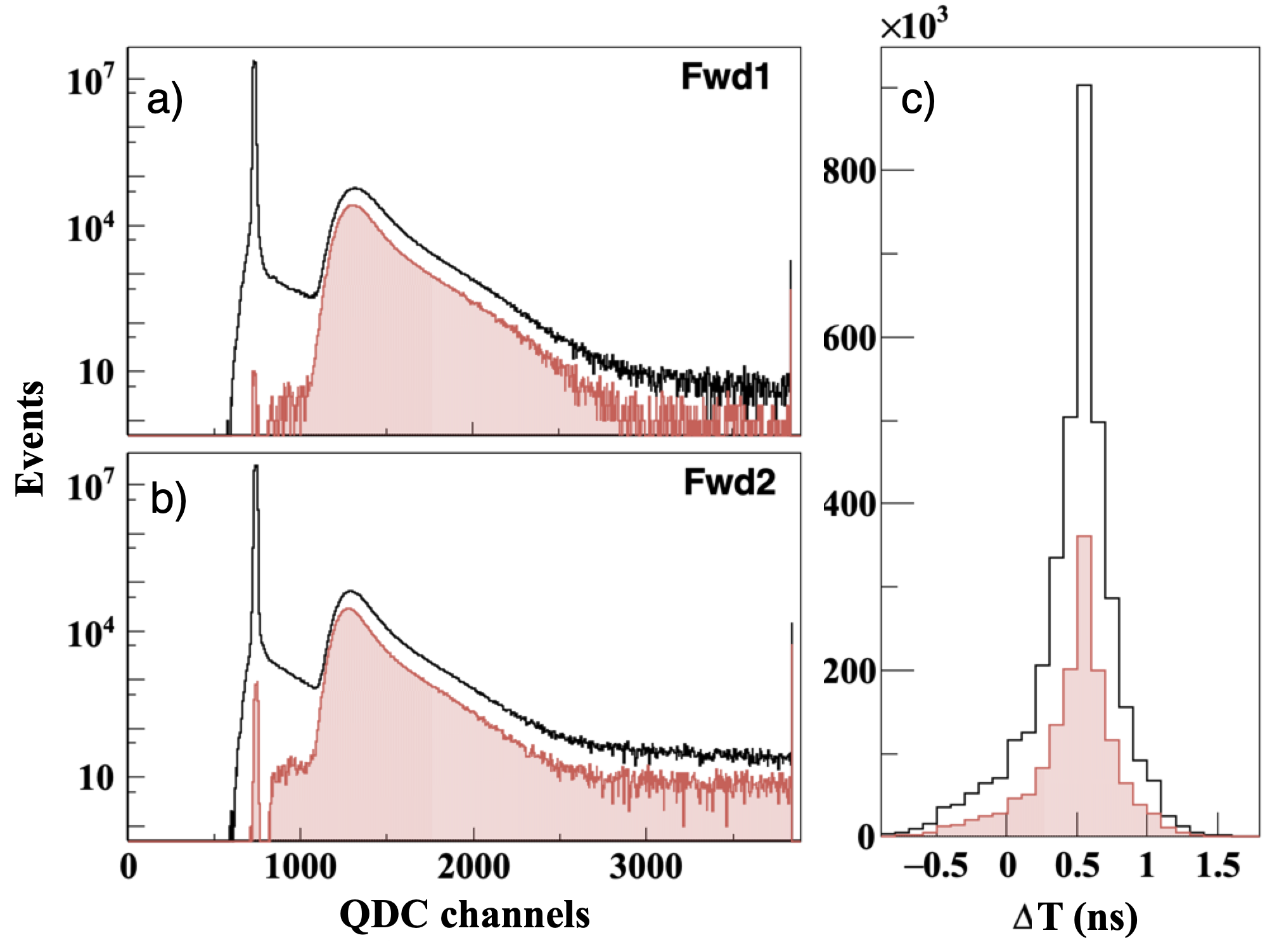}
  \caption{Response of the forward detector: a) the QDC spectrum of the first forward detector module on +X axis;
    b) the QDC spectrum of the second forward detector module on +X axis; c) Time difference between these two modules.
    The black curves are from the events which are triggered by the recoil
    detector.
    The red shaded curves are from the elastic scattering event sample.
  }
  \label{fig:fwd_performance}
\end{figure}

Typical QDC spectra from the two detector modules on the +X axis are shown as
black curves in Fig. \ref{fig:fwd_performance} (a) and (b).
They are obtained by selecting the events which generate a trigger in the recoil detector.
The peaks at around \num{1200} QDC channels are from the beam particles traversing through the scintillators, which have the same energy deposit as MIPs.
The pedestal noise peak is at about \num{735} QDC channels with a width of
$\sim\SI{6}{channels}$ ($\sigma$).
The timing resolution (FWHM) of the forward detector is about \SI{400}{ps}, which is determined from the spread of the time difference between the two modules as shown in Fig. \ref{fig:fwd_performance} (c).

A further study is carried out using the tag-and-probe method with a controlled
sample of elastic scattering events.
The response of one module is probed by using the other module to tag the sample
(the procedures for selecting the elastic scattering events are described in Sec. \ref{sec:tofe_selection}).
The results are shown as the red shaded curves in Fig. \ref{fig:fwd_performance}.
The lower limit of the elastic peaks is clearly seen at around \num{1000}.
In the following analysis, this value is used as the threshold for the forward detector.
The corresponding SNRs are about \num{40} for both detector modules.

The detection efficiencies of the forward modules are also determined with the
tag-and-probe method, which are about \SI{99.9978}{\percent} and \SI{99.9560}{\percent} respectively.
Scattering effects in the first layer reduce the detection efficiency slightly in the second layer.
The coincidence of both modules is required to have a valid hit in the
forward detector and the hit time is the average of two modules.

\subsection{Performance of the recoil detector}
\label{sec:recoil_performance}
\begin{figure}[b!]
  \centering
  \includegraphics[width=0.48\textwidth]{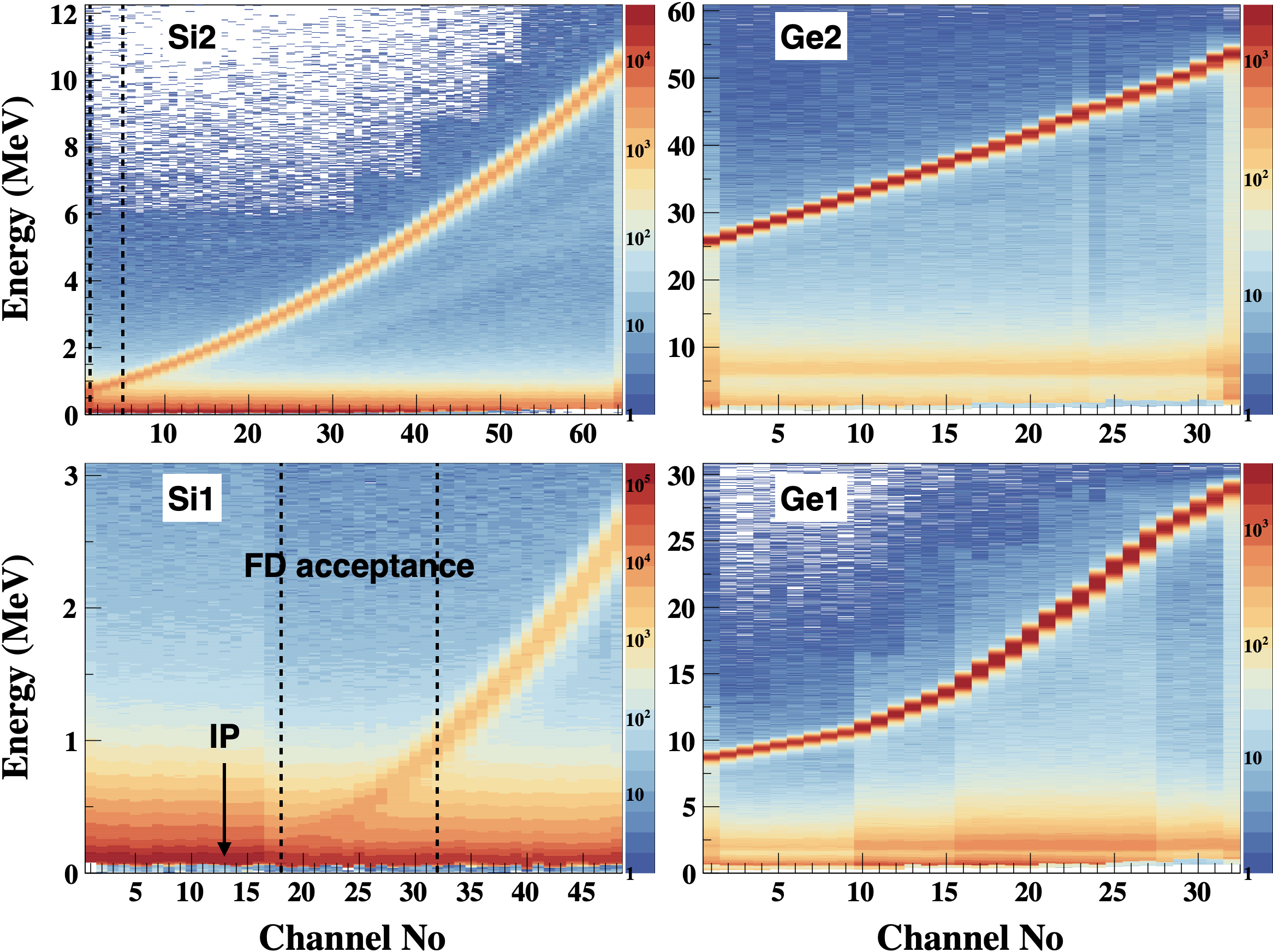}
  \caption{Energy spectra (after clustering) for all channels of the four recoil sensors obtained at \SI{2.2}{\momentum}.
    Smaller channel numbers indicate smaller recoil angle.
    IP indicates the channel which is aligned with
    the beam-target center.
    The group of channels which are expected to be fully covered by the forward detector are
    also indicated by the dashed lines on Si1 and Si2.
  }
  \label{fig:e_map}
\end{figure}

The reconstructed energy spectra of the recoil detector show a clear pattern
of elastic scattering in the distribution of the deposited energy versus the
position along the beam axis as shown in Fig. \ref{fig:e_map}.
The elastic peak is on a wide background, which exists over the whole energy range.
Although the elastic peak is well separated from the low-energy background at
large recoil angles, it becomes hard to distinguish at small recoil angles.
The background consists of three main components:
1) a fast-decreasing exponential component, which has very high yield at low energy;
2) a slow-decreasing exponential component, which extends to high energy;
3) a MIP component, which can be described by a Landau distribution and has a most probable energy deposit of \SI{0.37}{\MeV} in Si1 and Si2, \SI{2.2}{\MeV} in Ge1 and \SI{7.0}{\MeV} in Ge2.
The MIP events are mainly generated by charged pions produced by inelastic events.

\begin{figure}[tb!]
  \centering
  \includegraphics[width=0.45\textwidth]{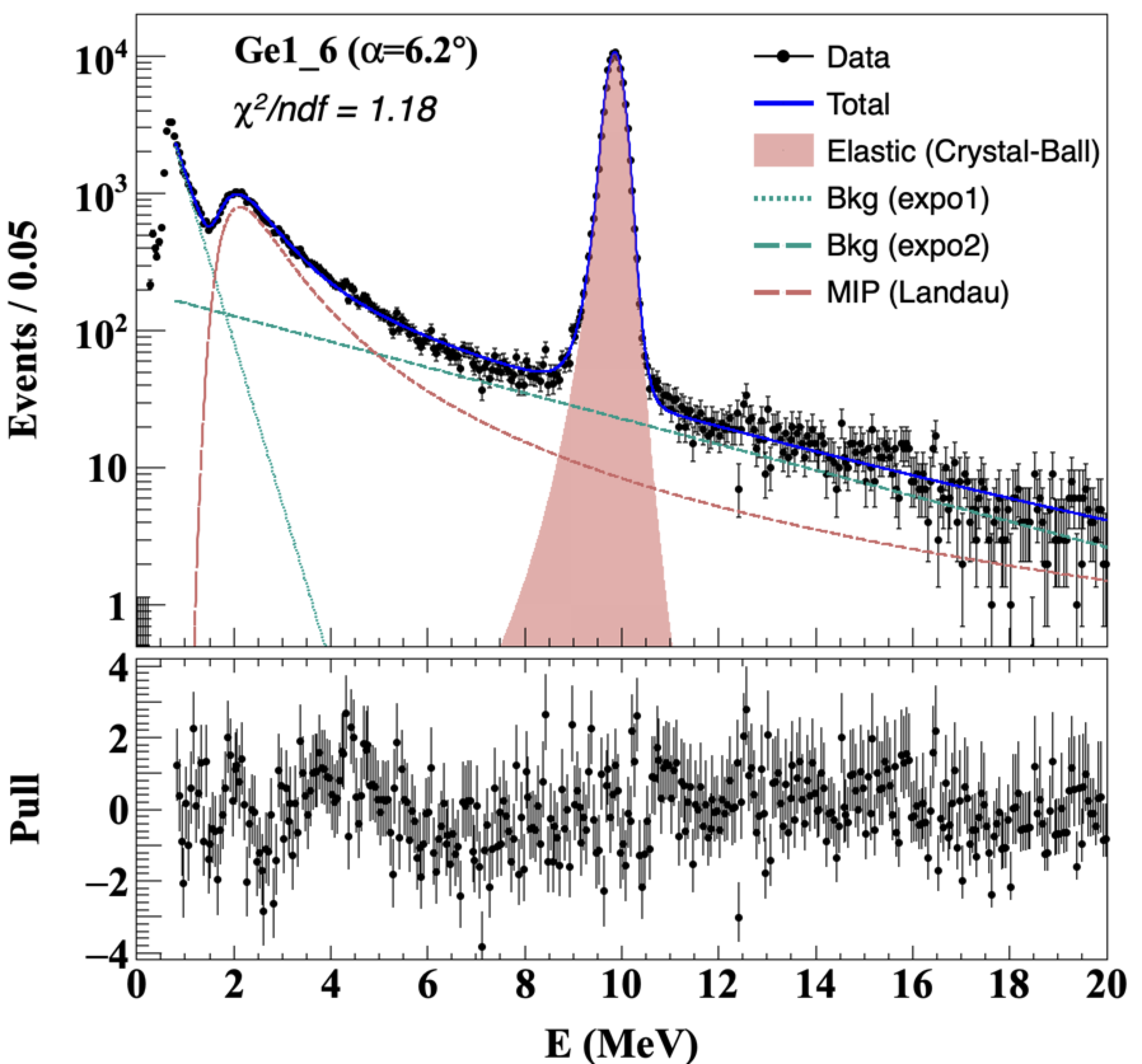}
  \caption{Extraction of the elastic peak spectrum using the extended binned
    Maximum-Likelihood fit. Three components are used in the background model
    and the fit is carried out over the full range indicated. The pull of each bin
    is shown in the bottom frame.}
  \label{fig:e_fit}
\end{figure}

The response of each strip to the elastic scattering events from the IP is described
very well by the Crystal-Ball function\cite{crystal_ball}, which is composed of a Gaussian core and a power-law tail on each side of the core.
For the strips at large recoil angle, an accurate estimate of the parameters of
the background components is possible by fitting the sidebands around the elastic peak.
An extended binned Maximum-Likelihood fit over the full energy range above the
trigger threshold, including both the background and the
elastic peak, is then carried out to extract the parameters of the response function and the event rate under the peak.
An example of the combined fit is shown in Fig. \ref{fig:e_fit}.
The sum of multiple Crystal-Ball functions, which share the same shape parameters and
different Gaussian peaks, are used to describe the response of the readout channel with multi-strips. 
The intrinsic resolution of the recoil sensor determined from the $\alpha$ source
calibration ($\sim$\SI{20}{\keV} FWHM at \SI{5}{\MeV}) is much smaller than the
extracted width of the elastic peak from the beam data ($\sim$\SI{330}{\keV}
FWHM at \SI{5}{\MeV}).
Thus, the width of the energy peak is mainly dominated by the finite
angular coverage of the strip and the thickness of the cluster jet target along the beam direction.

The accuracy of the combined fit deterioates at small recoil angles, when the elastic peak approaches the MIP peak and both peaks are embedded among the large
low-energy background ($\lesssim$ \SI{1}{\MeV}).
In this case, the fit parameters become strongly correlated and there is a large error in determining the fraction of each component.
A pre-selection of the elastic scattering events is needed to get a cleaner spectrum.
For the strips which are covered by the forward detector, the coincidence between the recoil proton and the scattering beam particle is utilized for this selection.

\subsection{Event selection with the TOF-E relation}
\label{sec:tofe_selection}

Most background events are already suppressed by requiring the forward detector to have a valid hit from the beam particle.
A more accurate selection of elastic scattering events is achieved by using the TOF-E relation.
The time-of-flight (TOF) of the recoil proton and its kinetic energy (E)
have a fixed relation $TOF = l\sqrt{m_p/2E}$, where $l$ is the distance of
the recoil detector to the IP and $m_p$ is the proton mass.
Due to the negligibly small variation of the flight time of the scattering beam particle,
the TOF of the recoil proton can be approximated by the difference between the hit time of the recoil detector and the forward detector.

\begin{figure}[b!]
  \centering
  \includegraphics[width=0.42\textwidth]{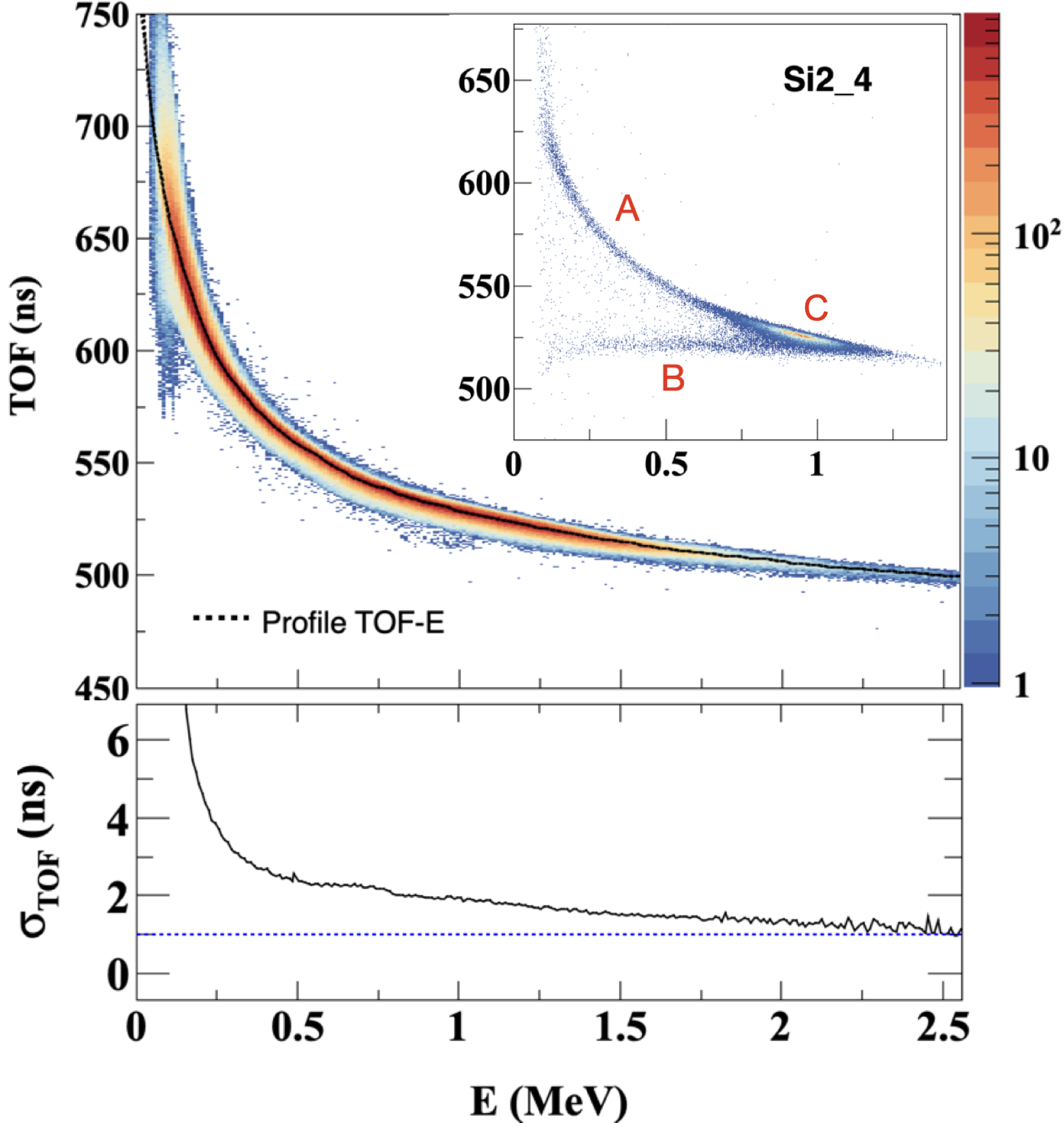}
  \caption{
    The upper frame shows the TOF-E spectrum of all strips for selected elastic
    scattering events at \SI{2.2}{\momentum}.
    A typical raw TOF-E spectrum for a single strip is shown in the inset, here region C is the elastic peak.
    The lower frame shows the time resolution for $\SI{10}{\keV}$ bins of energy.
    }
  \label{fig:tof-e}
\end{figure}

Two bands of events, which connect to the elastic peak, are observed in the raw TOF-E spectrum for each strip, as shown in the inset of Fig. \ref{fig:tof-e}.
Both bands are generated by the recoil protons of elastic scattering, with:
\begin{enumerate*}[label=(\roman*)]
\item band A from the interaction of the beam particle with the residual gas in the scattering chamber;
\item band B from the events in which the recoil protons hit the edge of the strip and only deposit part of their energy in the sensor.
\end{enumerate*}
The events of the elastic peak (region C in Fig. \ref{fig:tof-e}) and band A are selected roughly and filled into an overall TOF-E spectrum of elastic scattering events from all strips.
This spectrum is divided into \SI{10}{\keV} steps,  and a smooth, data-deduced TOF-E curve is obtained as shown in the main plot of Fig. \ref{fig:tof-e}.
The standard deviation of TOF ($\sigma_{TOF}$) for each step is also extracted and shown in the bottom frame of Fig. \ref{fig:tof-e}.
$\sigma_{TOF}$ reaches an asymptotic value of $\sim$\SI{1}{\ns} for high energies, which can be considered as the intrinsic timing resolution of the recoil
detector since the forward detector has much better resolution (see Sec. \ref{sec:fwd_performance}).
At $E < \SI{250}{\keV}$, $\sigma_{TOF}$ deterioates rapidly because of the
combined effect of the finite energy resolution and the large derivative of the TOF-E relation.

\begin{figure}[b!]
  \centering
  \includegraphics[width=0.45\textwidth]{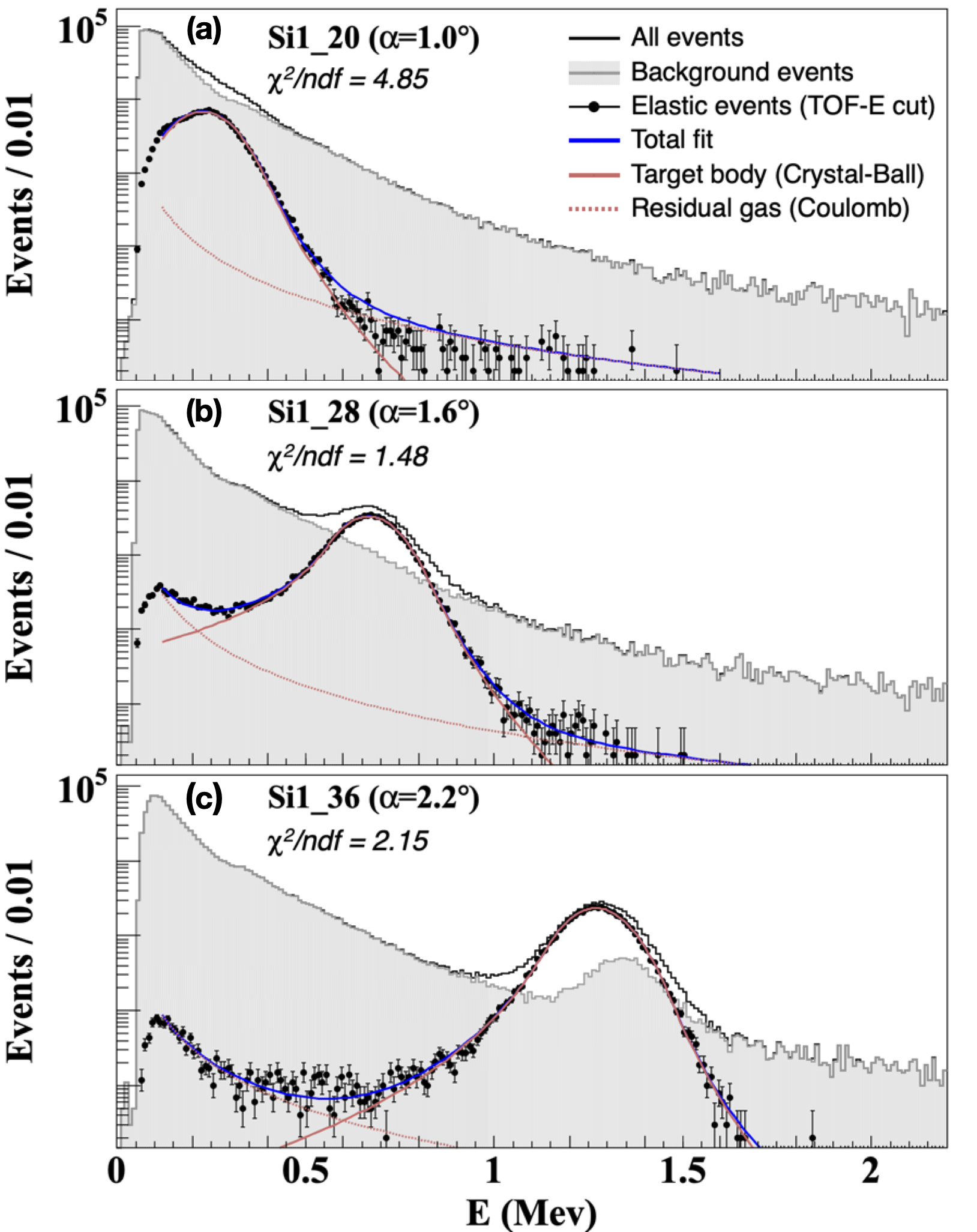}
  \caption{The energy spectra of TOF-E selected events (black dots) from strips at three recoil angles. The black lines are the spectra from all events and the grey spectra are from the events rejected since they had no coincidence with the forward detector.}
  \label{fig:cut}
\end{figure}

Events within $\pm 5\sigma_{TOF}$ of the TOF-E curve are selected as the elastic
scattering events.
A clean energy spectrum is obtained after the selection and the elastic peak shows up clearly.
The results for three strips at different recoil angles are shown in Fig. \ref{fig:cut}.
For strips which are fully covered by the forward detector, the spectrum of the
suppressed background events has a smooth transition over the elastic peak range
as shown in Fig. \ref{fig:cut} (b).
This indicates a high efficiency of the selection condition.
With increasing recoil angle, the forward detector gradually loses the full coverage
of the recoil strip and a small elastic peak appears in the background spectrum
as shown in Fig. \ref{fig:cut} (c).
On the other hand, with decreasing recoil angle, the finite thickness of the
cluster jet target can not be ignored and the forward detector starts to lose the coverage of the full target profile due
to the lower limit of the acceptance as shown in Fig. \ref{fig:cut} (a).
In comparison to Fig. \ref{fig:forward_acceptance}, the strips with all events
inside the red box are shown here in Fig. \ref{fig:cut} (b).
The strips where some of the events are lost above or below the red box are
shown in Fig. \ref{fig:cut} (c).
Strips with a significant fraction of events located near the vertical red line
at $X = \SI{3}{\cm}$ are shown in Fig. \ref{fig:cut} (a).

The remaining background in the selected events comes from elastic scattering
off the residual gas, which is mainly the hydrogen atoms from the evaporation
off the cluster jet beam \cite{cluster_target}.
The shape of this background is well described by the Coulomb elastic scattering cross
section, due to the uniform distribution of the residual gas and the rapid decrease of the elastic scattering cross section beyond the Coulomb region.
Again, an extended unbinned Maximum-Likelihood fit with a sum of the Coulomb elastic scattering formula and the Crystal-Ball
function is used to extract the parameters of the response function and the
yield of elastic events from the core of the cluster jet stream. 
Fit results for the three example strips are also shown in Fig. \ref{fig:cut}.
It's found that the fit quality deterioates when the full shape of the target profile is not available.
This limits the lower range of the $|t|$ measurement to $\sim$\SI{0.001}{\tmom} at $P_{beam}=\SI{2.2}{\momentum}$.

\subsection{Minimum measurable $|t|$}
\label{sec:minimum_t}
\begin{figure}[b!]
  \centering
  \includegraphics[width=0.45\textwidth]{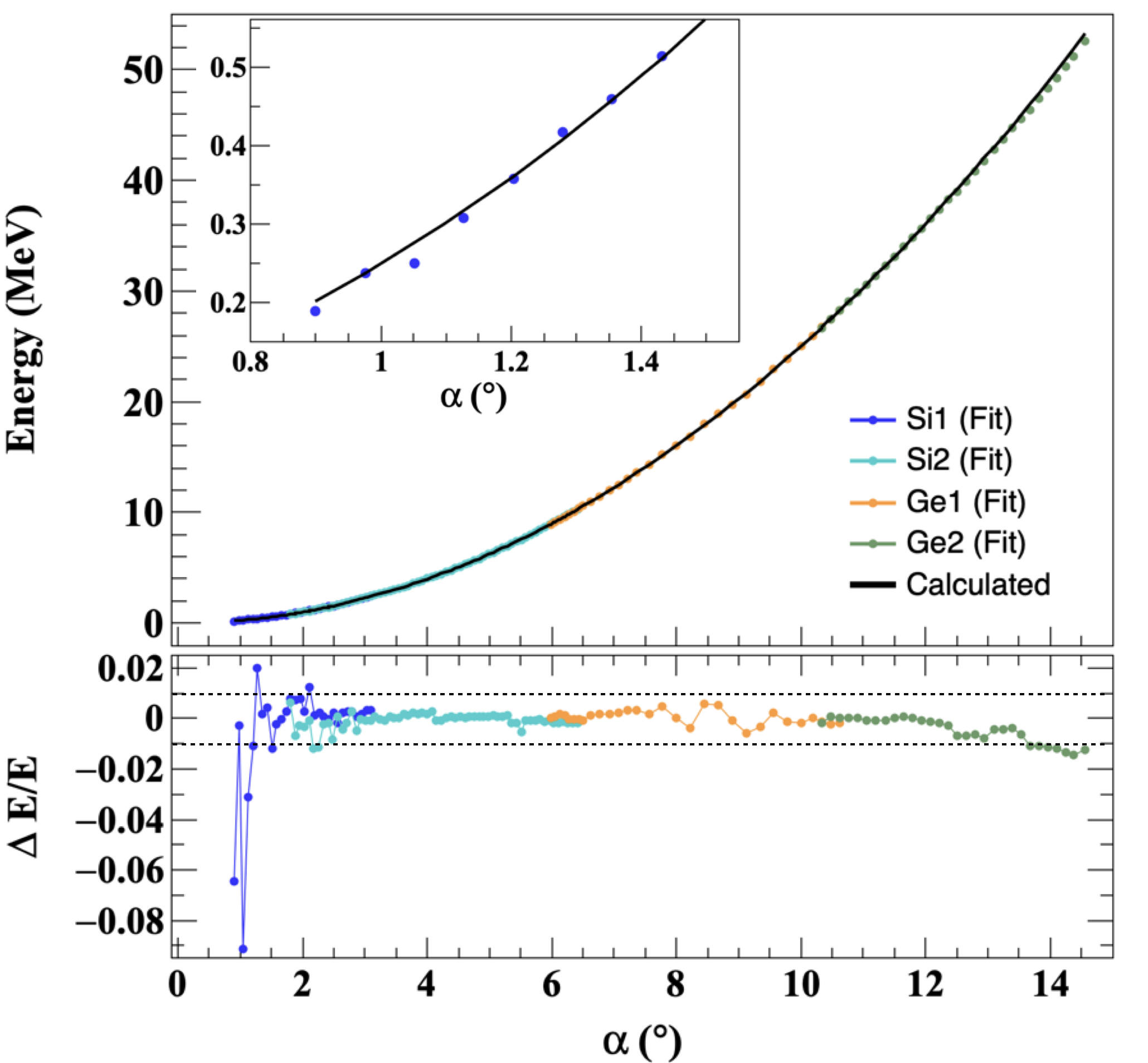}
  \caption{
    Comparison of the measured (colored dots) and the calculated (black line)
    recoil energy on the strip center at different recoil angles. The relative
    difference is drawn in the bottom plot. The results are for $P_{beam} = \SI{2.2}{\momentum}$.}
  \label{fig:measured_vs_calculated}
\end{figure}
The spectra of the suppressed background events on the strips which are fully covered by the
forward detector can be used as the template model to describe the background
shape for strips at smaller recoil angles.
Thus, the lower range of the $|t|$ measurement can be extended below the limit set by the TOF-E selection method.
The scaling parameter of the background model is first estimated with the upper
sideband of the energy spectrum, and a combined fit with the sum of the Crystal-Ball function and the background template is then carried
out over the full energy range to extract the elastic peak spectrum.

The limit of the method is reflected in the extracted elastic peak energy on each strip as shown in
Fig. \ref{fig:measured_vs_calculated}.
For the strips where the full elastic peak spectrum is extracted and described accurately
by the Crystal-Ball function, the relative difference between the measured and
the calculated recoil energy based on kinematic relations is within \SI{\pm 1}{\percent}.
The discrepancy enlarges rapidly at $\alpha < \SI{1.2}{\degree}$ when part of
the target profile is not recorded by the recoil strips due to the trigger threshold.
Both the extracted parameters of the response function and the event rate have large uncertainties in this case.
The observed lower limit of the energy peaks is about \SI{350}{\keV} at \SI{2.2}{\momentum}, which corresponds to the minimum measurable $|t|$ of $\sim$\SI{0.0007}{\tmom}.

\section{Conclusion and Outlook}
\label{sec:conclusion}

The commissioning of the full KOALA setup at COSY, specifically of the new forward detector, was successful.
A strong correlation between the recoil detector and the forward detector is
observed for the covered region of the forward detector.
The discriminating power of using the TOF-E relation of the recoil proton to select elastic
scattering events is verified.
Preliminary analysis shows that the lower |t| range can be extended slightly beyond the design goal of \SI{0.0008}{\tmom}.

The finite thickness of the target profile can not be ignored at
very small recoil angles ($\alpha < \SI{1.2}{\degree}$) and it greatly
constrains the lower limit of the $|t|$ measurement.
The observed integral thickness of the target profile is larger than the design value.
Since the stochastic cooling was not stable during the beam commissioning, this may be caused by the combined effect of the tilting of the cluster jet
beam and the unexpected large emittance of the COSY beam.
An investigation of the cluster target setup is ongoing and more stable stochastic cooling is required in the future experiments.
Furthermore, a larger size of the forward scintillators, especially a larger width, is also proposed.

The DAQ was operated in a stable condition during the beam commissioning.
However, bias of the trigger efficiency is observed among different sub-detectors.
This is caused by the electronic noise level in the COSY environment
and the limited performance of the mask gate based event synchronization.
Due to the self-triggering design, different noise triggering levels will change the corresponding trigger efficiency.
Thus, the mask gate in the trigger logic is proposed to be removed and the timestamp-based synchronization to be used in the future experiment.
A better method for the noise suppression is also under investigation.

\end{document}